\documentclass[pre,aps,nofootinbib]{sapm}

\usepackage{amsmath}
\usepackage{graphicx}
\usepackage{amssymb}
\usepackage{color}

\def\jcm#1{{\color{blue} #1}}

\begin{document}

\title{Nonlinear instabilities of multi-site breathers in Klein--Gordon lattices}

\author[J. Cuevas--Maraver, P.G. Kevrekidis and D.E. Pelinovsky]{Jes\'us Cuevas--Maraver, Panayotis G. Kevrekidis and Dmitry E. Pelinovsky}
\affil{Grupo de F\'{\i}sica No Lineal, Departamento de F\'{\i}sica Aplicada I, Universidad de Sevilla. Escuela Polit{\'e}cnica Superior, C/ Virgen de \'Africa, 7, 41011-Sevilla, Spain;\\ Instituto de Matem\'aticas de la Universidad de Sevilla (IMUS). Edificio Celestino Mutis. Avda. Reina Mercedes s/n, 41012-Sevilla, Spain}

\affil{Department of Mathematics and Statistics, University of
Massachusetts, Amherst, MA 01003-9305, USA;\\ Center for Nonlinear Studies and Theoretical Division, Los Alamos National Laboratory, Los Alamos, New Mexico 87545, USA}

\affil{Department of Mathematics, McMaster University, Hamilton, Ontario, Canada, L8S 4K1;\\ Department of Applied Mathematics, Nizhny Novgorod State Technical University, Nizhny Novgorod, Russia}

\maketitle

\begin{abstract}
In the present work, we explore the possibility of excited breather states
in a nonlinear Klein--Gordon lattice to become nonlinearly unstable, {\it even
if} they are found to be spectrally stable. The mechanism for this
fundamentally nonlinear instability is through the resonance
with the wave continuum of a multiple
of an internal mode eigenfrequency in the linearization of excited breather states. For the nonlinear instability, the internal mode must have
its Krein signature opposite to that of the wave continuum. This mechanism is not only theoretically proposed,
but also numerically corroborated through two concrete examples of the Klein--Gordon lattice
with a soft (Morse) and a hard ($\phi^4$) potential.
Compared to the case of the nonlinear Schr{\"o}dinger lattice, the Krein
signature of the internal mode relative to that of the wave continuum may change depending on
the period of the excited breather state. For the periods
for which the Krein signatures
of the internal mode and the wave continuum coincide, excited breather states are observed
to be nonlinearly stable.
%A generalization to vortices of the Klein--Gordon lattice in two spatial dimensions
%is also considered.
\end{abstract}

\section{Introduction}

The study of anharmonic modes  constructed out of a few
excited lattices sites in nonlinear lattice dynamical
systems is a broad and diverse theme of research
that emerged in the physics literature
through the work of~\cite{pa90,sietak}. Subsequently, the mathematical
proof of the existence of such modes in~\cite{MA94} not only placed
such states on a rigorous mathematical footing, but also gave a
deep set of insights towards their potentially generic nature.
This, in turn, led to a considerable growth and diversification
of interest into these modes over the last two decades, eventually
culminating in a wide array of reviews on both the methods of
analysis of these modes, as well as on their diverse
applications~\cite{Aubry,flach1,flach2,macrev}. Noting only
some among the many areas where these modes have been
impacting, we cite coupled waveguide arrays and photorefractive crystals
in the realm of nonlinear optics \cite{photon,review_opt},
the denaturation dynamics of the DNA double strand in
biophysics~\cite{peyrard}, breather formation
in eletrical lattices and in
micromechanical cantilever arrays~\cite{sievers}, Bose-Einstein
condensates in optical
lattices in atomic physics \cite{morsch}, as well as bright and
dark breathers in granular crystals~\cite{theo,chong}.

Orbital and asymptotic stability of the fundamental (single-site) breathers were established by Bambusi \cite{Bambusi1,Bambusi2}.
Spectral stability of more complicated multi-breathers were classified in the recent works \cite{Archilla,KK09,PelSak},
depending on the phase difference in the nonlinear oscillations between different sites of the lattice (and the nature of the potential).
The simplest spectrally stable multi-breather configuration includes a two-site breather.
If the two sites excited in the anti-continuum limit are adjacent in the lattice, then the spectrally stable breather
is in-phase (anti-phase) for the hard (soft) potential $V$ \cite{KK09,PelSak}.

The main question we would like to consider is if the spectrally stable two-site breather
is also stable in the nonlinear dynamics of the discrete Klein--Gordon (KG) equation (\ref{eq:dyn1d}).
This is a part of the more general and broadly important
question about whether spectrally stable excited (non-fundamental) states
of a Hamiltonian system are nonlinearly stable or not.
In a similar context of the discrete nonlinear Schr\"{o}dinger
(NLS) equation, the two-site breathers near the anti-continuum limit
are not orbitally stable because the linearized spectrum
admits an internal mode of negative energy (negative Krein signature) \cite{PKF1}.
This internal mode of negative energy destabilizes nonlinear
dynamics of the two-site breathers if a multiple harmonic of the internal mode eigenfrequency
occurs in the frequency spectrum of the wave continuum that bears positive energy \cite{Cuccagna,KPS}.

In the context of the discrete KG equation (\ref{eq:dyn1d}), a similar concept of
Krein signature can be introduced for the internal modes in the linearization of multi-site
breather solutions \cite{Aubry,MS98,MAF98}. The discrete NLS equation appears to be
a valid approximation for the small-amplitude weakly coupled multi-site breathers
\cite{PPP}. Therefore, it is natural to expect that the
mechanism of nonlinear instability based on the resonance between multiple harmonics of
the internal mode eigenfrequency and the frequency spectrum of the wave continuum may also be observed for the two-site breathers
near the anti-continuum limit. This conjecture is fully confirmed in this work.
However, depending on the period of the limiting breather, one can find
parameter configurations, where the nonlinear instability can be avoided, because
the Krein signatures of both the internal mode and the wave continuum coincide. In such situations,
our analytical and numerical results indicate that the two-site breathers are
stable in the nonlinear dynamics of the discrete KG equation (\ref{eq:dyn1d}).

The presentation of our results is structured as follows.
In section 2, we present the mathematical formalism of the
problem. Upon setting up the Klein-Gordon lattices, their
breather solutions and associated linearization (consisting of
both internal modes and wave continuum), we explore the notion of
their Krein signature. Subsequently, we use asymptotic
expansions to reveal the nonlinear resonant mechanism leading to the
main instability result of the present work. In section 3, we numerically
corroborate the analytical results. We report computations of
the Floquet spectrum associated with the breathers in two examples of
the Klein--Gordon lattice associated with the Morse and hard $\phi^4$ potentials.
We also present direct numerical
simulations highlighting the outcomes of the unstable or stable dynamics of two-site breathers.
%A generalization to vortex solutions in the discrete KG equation in two spatial dimensions is described
%in section IV.
Finally, in section 4 we summarize our results and present some
directions for future work.

\section{Mathematical formalism}

We start with the one-dimensional Klein--Gordon (KG) lattice equation:
\begin{eqnarray}\label{eq:dyn1d}
    \ddot u_n+V'(u_n) = \varepsilon(u_{n+1} - 2u_n + u_{n-1}), \quad n \in \mathbb{Z},
\end{eqnarray}
where $V : \mathbb{R} \to \mathbb{R}$ is an on-site (substrate) potential and $\varepsilon > 0$ is the coupling constant.
The amplitudes of coupled nonlinear oscillators on lattice sites form a sequence $\{ u_n \}_{n \in \mathbb{Z}} \in \mathbb{R}^{\mathbb{Z}}$
denoted by the vector ${\bf u}$, which is typically defined in the sequence space $\ell^2(\mathbb{Z})$.
The solution ${\bf u}$ of the discrete KG equation (\ref{eq:dyn1d}) is a function of time $t$. A local
solution ${\bf u}$ exists in $C^1((-T,T);\ell^2(\mathbb{Z}))$ for some $T > 0$ if $V'$ is Lipschitz,
thanks to the Picard's contraction method and the boundedness of the discrete Laplace operator
in the lattice equation (\ref{eq:dyn1d}).

\begin{figure}
\begin{center}
\begin{tabular}{cc}
\includegraphics[width=5.5cm]{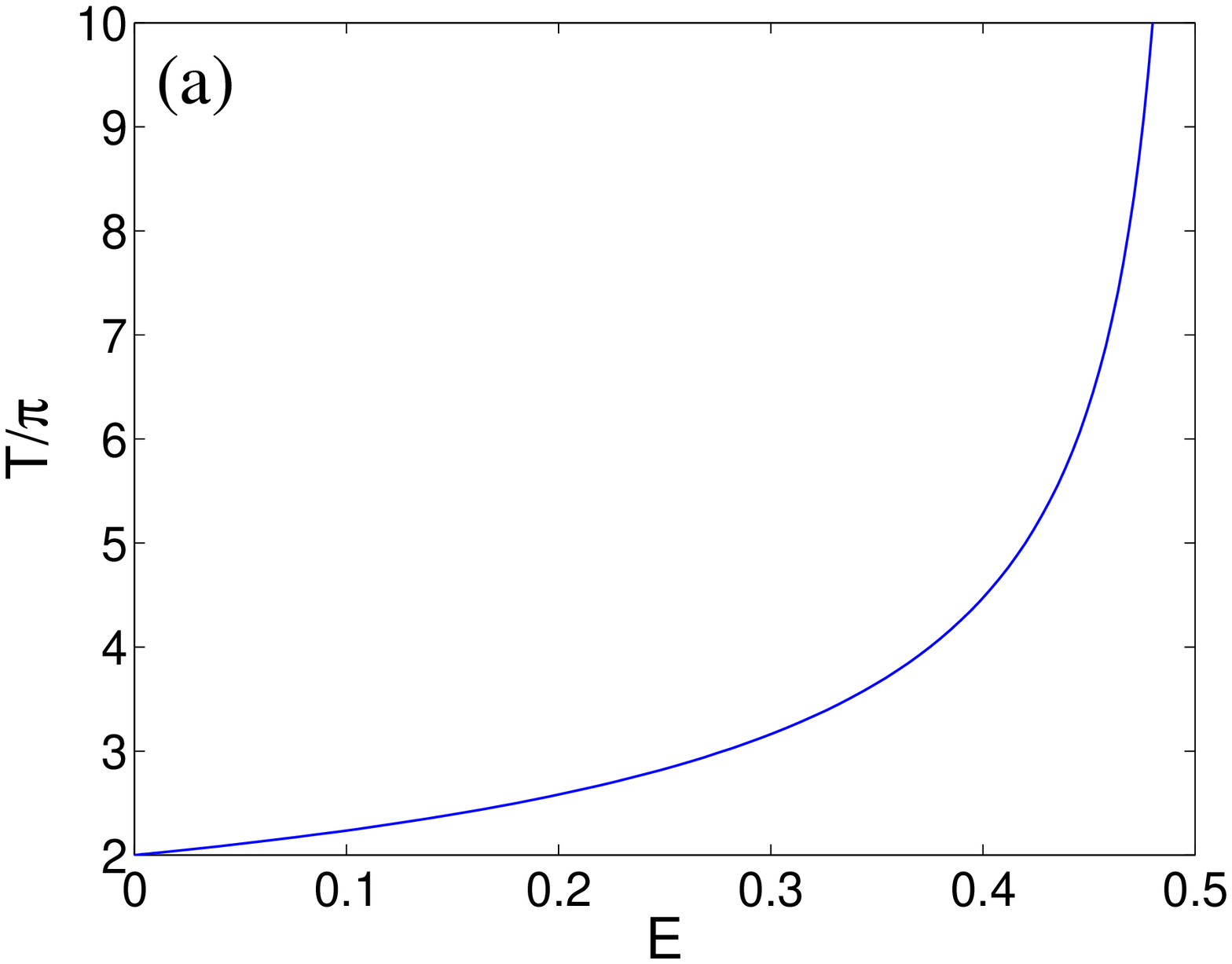} &
\includegraphics[width=5.5cm]{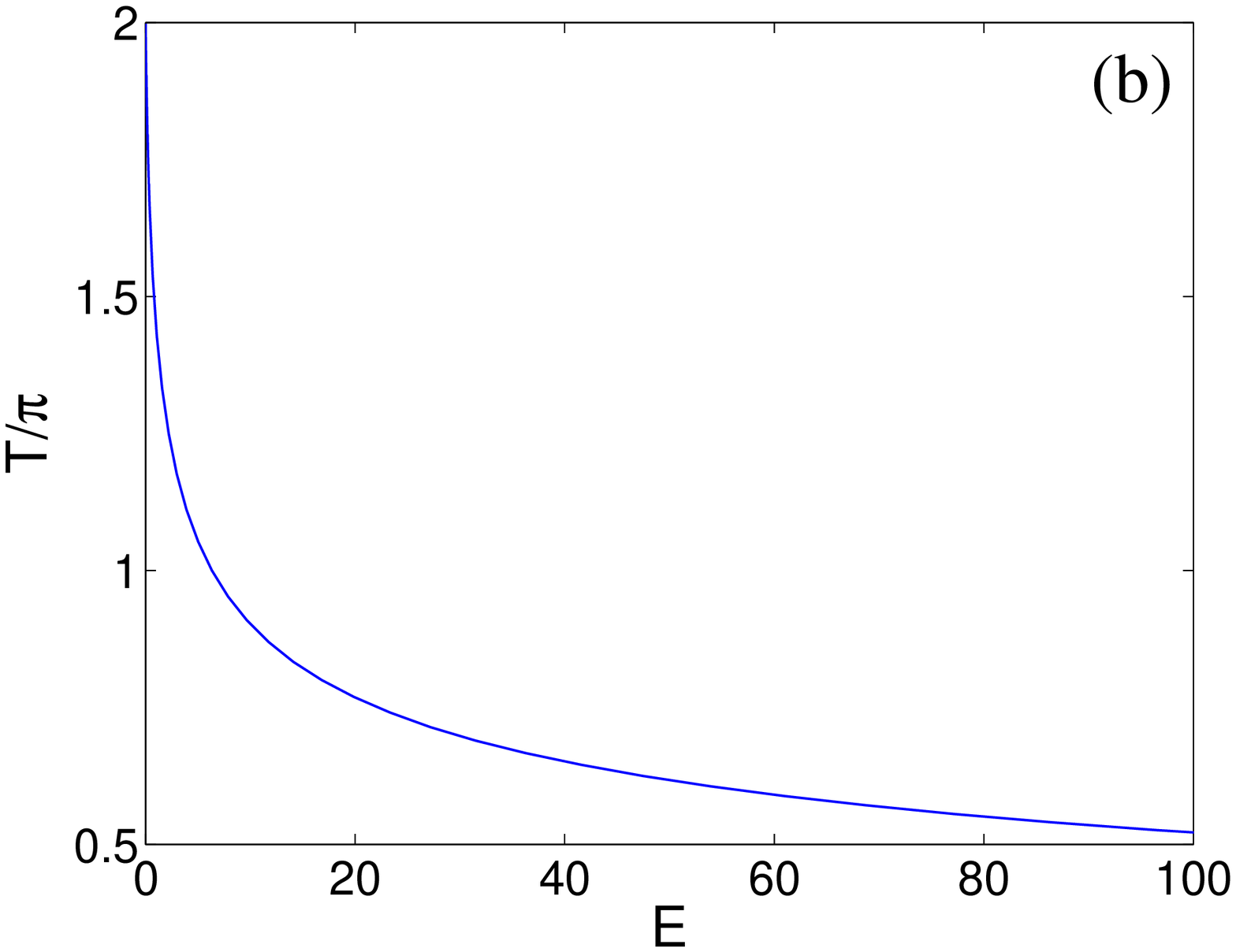} \\
\end{tabular}
\end{center}
\caption{Dependence of the period $T$ with respect to the energy $E$
for the nonlinear oscillator (\ref{oscillator}) in the Morse (a) and
$\phi^4$ (b) potentials.}
\label{fig:periods}
\end{figure}

For numerical experiments, we consider two prototypical
examples of smooth on-site potentials, namely the Morse
and $\phi^4$ potentials, which are given, respectively, by
\begin{equation}
\label{potentials}
\mbox{\rm (i)} \;\; V(u) = \frac{1}{2} (e^{-u} - 1)^2 \quad \mbox{\rm and}\quad
\mbox{\rm (ii)} \;\; V(u) = \frac{1}{2} u^2 + \frac{1}{4} u^4.
\end{equation}
We note that $V(0) = V'(0) = 0$ and $V''(0) = 1$ due to our normalization.
The Morse potential (i) in (\ref{potentials}) is classified as a
soft potential
because the period $T$ of oscillations in the nonlinear oscillator equation
\begin{equation}
\label{oscillator}
\ddot{\varphi} + V'(\varphi) = 0
\end{equation}
increases with the oscillator energy $E = \frac{1}{2} \dot{\varphi}^2 + V(\varphi)$ \cite{CKKA11}.
The $\phi^4$ potential (ii) in (\ref{potentials}) is a hard potential because the period $T$ decreases with $E$.
Fig. \ref{fig:periods} shows the dependence of the period $T$ with respect to the energy $E$
for the Morse (a) and $\phi^4$ (b) potentials.

The discrete KG equation (\ref{eq:dyn1d}) is associated with the Hamiltonian
\begin{eqnarray}\label{eq:ham1d}
    H = \frac{1}{2} \sum_{n \in \mathbb{Z}} \dot{u}_n^2 + \frac{\varepsilon}{2} \sum_{n \in \mathbb{Z}} (u_{n+1} - u_n)^2 + \sum_{n \in \mathbb{Z}} V(u_n).
\end{eqnarray}
The energy conservation (\ref{eq:ham1d}) can be used to extend the local solution ${\bf u} \in C^1((-T,T);\ell^2(\mathbb{Z}))$
to the global solution ${\bf u} \in C^1(\mathbb{R};\ell^2(\mathbb{Z}))$ if the potential $V$
possesses some coercivity. This is definitely the case for the hard $\phi^4$ potential
(ii) in (\ref{potentials}), which supports global solutions of the KG lattice equation (\ref{eq:dyn1d})
with time-independent bounds on $\| {\bf u} \|_{\ell^2}$ and $\| \dot{\bf u} \|_{\ell^2}$.
Due to positivity of $V$ for the soft Morse potential (i),
the local solution ${\bf u} \in C^1((-T,T);\ell^2(\mathbb{Z}))$ is extended for $\varepsilon > 0$
to the global solution with time-uniform bounds on $\| \dot{\bf u} \|_{\ell^2}$
and $\| \delta {\bf u} \|_{\ell^2}$, where
$\delta$ is the difference operator defined by $(\delta {\bf u})_n = u_{n+1} - u_n$,
$n \in \mathbb{Z}$. By using $\frac{d}{dt} \| {\bf u} \|_{\ell^2} \leq \| \dot{\bf u} \|_{\ell^2}$,
we obtain a global but linearly growing bound on $\| {\bf u} \|_{\ell^2}$.

Let us mention in passing that the soft $\phi^4$ potential
$V(u) = \frac{1}{2} u^2 - \frac{1}{4} u^4$ supports the blow-up in a finite time
for sufficiently large initial data~\cite{karachal}.
To avoid the finite-time blow-up, we take the Morse potential $V$ in (i) as a representative
of the class of soft potentials. The Morse potential has also applications in physics of DNA~\cite{rapti}.

We consider the breather solutions of the discrete KG equation (\ref{eq:dyn1d}),
which are $T$-periodic in time and
exponentially localized in space. These solutions
are given by the vector ${\bf u}(t) = {\bf u}(t+T)$ such that
${\bf u} \in C^{\infty}_{\rm per}((0,T);\ell^2(\mathbb{Z}))$ if $V$ is smooth.
Breather solutions are constructed by the arguments based on the implicit function theorem
for sufficiently small $\varepsilon > 0$ \cite{MA94} (see also review in \cite{PelSak}).
In what follows, we always assume that the potential $V$ is smooth.

Adding a perturbation ${\bf w} \in C^{\infty}(\mathbb{R};\ell^2(\mathbb{Z}))$
to the breather solution ${\bf u}(t) = {\bf u}(t+T)$, where ${\bf u} \in C^{\infty}_{\rm per}((0,T);\ell^2(\mathbb{Z}))$,
we obtain the linearized discrete KG equation in the form
\begin{equation}
\label{linKG}
\ddot w_n + V''(u_n) w_n = \varepsilon(w_{n+1} - 2w_n + w_{n-1}), \quad n \in \mathbb{Z}.
\end{equation}
According to the Floquet theory, we are looking for solutions of the linearized equation (\ref{linKG})
in the form ${\bf w}(t) = e^{\lambda t} {\bf W}(t)$, where $\lambda \in \mathbb{C}$ is a spectral parameter
and ${\bf W} \in C^{\infty}_{\rm per}((0,T);\ell^2(\mathbb{Z}))$ is the eigenvector of the
spectral problem
\begin{equation}
\label{spectrumKG}
\ddot W_n + 2 \lambda \dot{W}_n + \lambda^2 W_n + V''(u_n) W_n = \varepsilon(W_{n+1} - 2W_n + W_{n-1}), \quad n \in \mathbb{Z}.
\end{equation}
Given an admissible value of the spectral parameter $\lambda$, we can compute the Floquet multiplier
$\mu$ by $\mu = e^{\lambda T}$. The breather ${\bf u}$ is spectrally stable if all admissible values
of the Floquet multipliers $\mu$ belong to the unit circle. Due to the reversibility and reality of the linearized discrete
KG equation (\ref{linKG}), the Floquet multipliers $\mu$ occur either in  pairs on the unit circle $\{ \mu, \bar{\mu} \}$
or real pairs $\{ \mu, \mu^{-1} \}$  or quartets off the unit circle $\{ \mu, \bar{\mu}, \mu^{-1}, \bar{\mu}^{-1} \}$.
The breathers are spectrally unstable if real pairs or complex quartets exist outside the unit circle.

\subsection{Wave continuum}

The wave continuum is defined by solutions of the limiting spectral problem
\begin{equation}
\label{spectrumKG-lim}
\ddot W_n + 2 \lambda \dot{W}_n + \lambda^2 W_n + W_n = \varepsilon(W_{n+1} - 2W_n + W_{n-1}), \quad n \in \mathbb{Z},
\end{equation}
which corresponds to the zero solution ${\bf u} = {\bf 0}$. Performing the discrete Fourier
transform,
\begin{equation}
\label{Fourier-transform}
W_n(t) = \int_{-\pi}^{\pi} \hat{W}(t,\theta) e^{i n \theta} d \theta, \quad n \in \mathbb{Z},
\end{equation}
and using the Fourier series for the $T$-periodic functions
\begin{equation}
\label{Fourier-series}
\hat{W}(t,\theta) = \sum_{m \in \mathbb{Z}} \hat{W}_m(\theta) e^{i m \omega_0 t}, \quad \omega_0 = \frac{2\pi}{T},
\end{equation}
we obtain the dispersion equation for the spectral bands
\begin{equation}
\label{spectral-bands}
\lambda = i (\pm \omega(\theta) - m \omega_0), \quad m \in \mathbb{Z},
\end{equation}
where the function $\omega : [-\pi,\pi] \to [1,\sqrt{1+4\varepsilon}]$ represents
the fundamental band of the wave continuum given by
\begin{equation}
\label{wave-spectrum}
\omega(\theta) := \sqrt{1+4\varepsilon \sin^2\left(\frac{\theta}{2}\right)}, \quad \theta \in [-\pi,\pi].
\end{equation}
In terms of the Floquet multiplier $\mu = e^{\lambda T}$, only two spectral bands
show up symmetrically on the unit circle.
The wave continuum corresponds to the Floquet multipliers at
$\mu = e^{\pm i \omega(\theta) T}$, $\theta \in [-\pi,\pi]$.
The two spectral bands shrink to the two points of infinite multiplicities at
$\mu = e^{\pm i T}$ in the anti-continuum limit $\varepsilon \to 0$.

\subsection{Krein signature}

The Krein quantity represents the symplectic structure
of the linearized discrete KG equation (\ref{linKG}). It is used to
characterize Floquet multipliers on the unit circle and
stability transitions of discrete breathers \cite{Aubry,MS98,MAF98}.
Recently, the same quantity is used in the context of stability of the periodic
travelling waves in the FPU lattices \cite{BP2013}.
Using variables $\{ w_n, p_n \}_{n \in \mathbb{Z}}$,
we can write (\ref{linKG}) in the form,
\begin{equation}
\label{linKG-symplectic}
\frac{d w_n}{d t} = \frac{\partial \mathcal{H}}{\partial p_n}, \quad
\frac{d p_n}{d t} = -\frac{\partial \mathcal{H}}{\partial w_n}, \quad n \in \mathbb{Z},
\end{equation}
where $\mathcal{H}$ is the second variation of the Hamiltonian $H$ in (\ref{eq:ham1d}) given by
\begin{eqnarray}\label{energy}
\mathcal{H} = \frac{1}{2} \sum_{n \in \mathbb{Z}} p_n^2 +
\frac{\varepsilon}{2} \sum_{n \in \mathbb{Z}} (w_{n+1} - w_n)^2 + \frac{1}{2} \sum_{n \in \mathbb{Z}} V''(u_n) w_n^2.
\end{eqnarray}
Then, the symplectic (Krein) quantity is given for any ${\bf w},{\bf p} \in \ell^2(\mathbb{Z})$ by
\begin{equation}
\label{Krein}
K = i \sum_{n \in \mathbb{Z}} (\bar{p}_n w_n - p_n \bar{w}_n).
\end{equation}
We note that $K$ is real and constant in time $t$. It is identically zero at the real-valued
solutions of the linearized discrete KG equation (\ref{linKG}).
However, it is nonzero for complex-valued solutions, e.g. for eigenfunctions
${\bf w}(t) = e^{i \Omega t} {\bf W}(t)$, where $\lambda = i \Omega$ and ${\bf W} \in C^{\infty}_{\rm per}((0,T);\ell^2(\mathbb{Z}))$
are solutions of the spectral problem (\ref{spectrumKG}). In this case, the Krein quantity $K$ can be written
in the equivalent form
\begin{equation}
\label{Krein-Omega}
K = 2 \Omega \sum_{n \in \mathbb{Z}} |W_n|^2 + i \sum_{n \in \mathbb{Z}} (\dot{\bar{W}}_n W_n - \dot{W}_n \bar{W}_n).
\end{equation}

For the wave continuum, given by the Fourier transform (\ref{Fourier-transform}) and the Fourier series (\ref{Fourier-series})
for $\Omega = \pm \omega(\theta) - m \omega_0$ with $\theta \in [-\pi,\pi]$ and $m \in \mathbb{Z}$,
we obtain
$$
K = \pm 4 \pi \int_{-\pi}^{\pi} \omega(\theta) |\hat{W}_m(\theta)|^2 d \theta.
$$
Since $\omega(\theta) > 0$ for every $\theta \in [-\pi,\pi]$, according to the dispersion relation
(\ref{wave-spectrum}), we have $K > 0$ for each $m$-th spectral band with $\Omega = \omega(\theta) - m \omega_0$
and $K < 0$ for each $m$-th spectral band with $\Omega = -\omega(\theta) - m \omega_0$, for every $m \in \mathbb{Z}$.

\subsection{Internal modes of multi-site breathers}

The internal modes correspond to isolated Floquet multipliers on the unit circle
that occur commonly in the linearization of multi-site breathers \cite{kivshar,pelkiv}.
The internal modes and their Krein signature are well approximated
near the anti-continuum limit of weak coupling between the nonlinear oscillators.
Asymptotic expansions for eigenvalues and eigenfunctions of the spectral problem (\ref{spectrumKG})
for small $\varepsilon$ are reported in the most general case in \cite{PelSak}.
Here we review these expansions in the particular case, when the nonlinear oscillators
are excited on adjacent lattice sites as $\varepsilon \to 0$ (see also \cite{CKKA11,KK09}).

Let us represent the multi-site breather by its limiting  configuration
\begin{equation}
\label{limiting-breather}
{\bf u}_{\rm lim}(t) := \lim_{\varepsilon \to 0} {\bf u}(t) = \sum_{j = 1}^N \varphi(t + (\sigma_j-1) T/4) {\bf e}_j,
\end{equation}
where $N$ is the number of excited oscillators, ${\bf e}_j$ is the unit vector supported at the $j$-th lattice node,
$\sigma_j \in \{+1,-1\}$ for $j = 1,2,...,N$, and
$\varphi \in C^{\infty}_{\rm per}(0,T)$ is a unique, even $T$-periodic solution of the nonlinear oscillator equation (\ref{oscillator})
such that $\varphi(0) > 0$. The two nonlinear oscillators at the $j$-th and $(j+1)$-th lattice nodes
are in phase if $\sigma_j \sigma_{j+1} = 1$ and anti-phase if $\sigma_j \sigma_{j+1} = -1$.
Under this choice of in-phase and anti-phase configurations, it is proved
that the multi-site breather solution ${\bf u} \in C^{\infty}_{\rm per}((0,T);\ell^2(\mathbb{Z}))$
is uniquely continued from its limit (\ref{limiting-breather}) for small $\varepsilon > 0$ and the correction term ${\bf u} - {\bf u}_{\rm lim}$
is $\mathcal{O}_{\ell^2}(\varepsilon)$ \cite{MA94,PelSak}.

The spectral stability problem (\ref{spectrumKG})  with $\varepsilon = 0$ associated with the limiting breather
(\ref{limiting-breather})  has the Floquet multiplier
$\mu = 1$ of algebraic multiplicity $2N$ and the pair of Floquet multipliers $\mu = e^{\pm i T}$ of
infinite algebraic multiplicities. The two spectral bands of the wave continuum on the unit circle bifurcate from the
pair of Floquet multipliers $\mu = e^{\pm i T}$. The Krein signature of the two spectral bands is ${\rm sign}(K) = \pm 1$.
Isolated Floquet multipliers may also bifurcate from the pair of Floquet multipliers $\mu = e^{\pm i T}$
to share the same Krein signature as the one for
the spectral band they bifurcate from. In the context of the discrete NLS equation,
the absence of isolated Floquet multipliers bifurcating from the spectral bands can be proved
under some restrictive conditions  \cite{PelSak2011}.
In either case, our main interest is to study internal modes arising as a result of
splitting of the Floquet multiplier
$\mu = 1$ of algebraic multiplicity $2N$ under the perturbation terms with small $\varepsilon$.

Representing the small spectral parameter $\lambda = \varepsilon^{\frac{1}{2}} \Lambda$ in the spectral
problem (\ref{spectrumKG}), the authors of \cite{PelSak} justified the following asymptotic expansion for
the eigenvector ${\bf W}$ of the spectral problem (\ref{spectrumKG}):
\begin{equation}
\label{asymptotic-expansion}
{\bf W}(t) = \sum_{j=1}^N c_j \dot{\varphi}(t  + (\sigma_j-1) T/4) {\bf e}_j -
\varepsilon^{\frac{1}{2}} \Lambda \sum_{j =1}^N c_j v(t + (\sigma_j-1) T/4) {\bf e}_j + \varepsilon {\bf W}_{\rm rem}(t),
\end{equation}
where the even $T$-periodic function $v \in C^{\infty}_{\rm per}(0,T)$ is uniquely determined by the solution
of the inhomogeneous equation
\begin{equation}
\label{inhomogeneous-eq}
\ddot{v} + V''(\varphi) v = 2 \ddot{\varphi},
\end{equation}
${\bf c} = (c_1,c_2,...,c_N)$ is the vector of projections to be determined,
$\Lambda$ is the new spectral parameter, and ${\bf W}_{\rm rem}$ is the remainder term of the
asymptotic expansion. The values of $\sigma_j \in \{+1,-1\}$ for $j = 1,2,...,N$
are the same as in (\ref{limiting-breather}).

In the case of symmetric potentials $V$ satisfying $V(-u) = V(u)$, it is shown in \cite{PelSak}
that ${\bf W}_{\rm rem} \in C^{\infty}_{\rm per}((0,T);\ell^2(\mathbb{Z}))$ exists only if
the spectral parameter $\Lambda$ is defined by
the matrix eigenvalue problem for the eigenvector ${\bf c}$:
\begin{equation}
\label{matrix-eigenvalue}
-\frac{T(E)^2}{S T'(E)} \Lambda^2 {\bf c} = M {\bf c},
\end{equation}
where $S = \int_0^T \dot{\varphi}^2(t) dt$ is a positive constant coefficient, $T'(E)$ is the derivative of the period
of the nonlinear oscillator equation (\ref{oscillator}) with respect to
its energy $E = \frac{1}{2} \dot{\varphi}^2 + V(\varphi)$, see Fig. \ref{fig:periods},
and $M$ is the tri-diagonal matrix given by the elements
$$
M_{ij} = \left\{ \begin{array}{l} - \sigma_j (\sigma_{j+1} + \sigma_{j-1}), \quad i = j, \\
\phantom{texttextt} 1, \phantom{texttextt} i = j \pm 1, \end{array} \right.
$$
subject to the Dirichlet boundary conditions. In the case of hard potential with $T'(E) < 0$
and for the two-site in-phase breather with $N = 2$ and $\sigma_1 = \sigma_2$, the matrix eigenvalue problem (\ref{matrix-eigenvalue})
has the double zero eigenvalue $\Lambda = 0$ and a pair of purely imaginary eigenvalues
\begin{equation}
\label{internal-mode}
\Lambda = \pm i \Omega_0, \quad \Omega_0 = \frac{\sqrt{2 S |T'(E)|}}{T(E)} > 0.
\end{equation}
The same conclusion holds for the soft potential with $T'(E) > 0$ but applies
to the two-site anti-phase breather with $N = 2$ and $\sigma_1 = -\sigma_2$.
The asymptotic approximation (\ref{internal-mode}) corresponds to the pair of Floquet
multipliers on the unit circle, which are given at the leading order by $\mu = e^{\pm i \varepsilon^{\frac{1}{2}} \Omega_0}$.

We shall now substitute the asymptotic expansion (\ref{asymptotic-expansion}) with
$\Lambda = i \Omega_0$ into the Krein quantity (\ref{Krein-Omega}) and compute
the Krein signature of the internal modes (\ref{internal-mode})
to the leading order in $\varepsilon$. After straightforward computations,
we obtain $K = \varepsilon^{\frac{1}{2}} K_1 + \mathcal{O}(\varepsilon)$, where
\begin{equation}
\label{Krein-first}
K_1 = 2 \Omega_0 \sum_{j=1}^N A |c_j|^2, \quad
A := \dot{\varphi}^2 - \dot{\varphi} \dot{v} + v \ddot{\varphi}.
\end{equation}
By the time-conservation of $K$, it is clear that $A$ is independent of $t$.
Let us compute this quantity explicitly by using the exact expression for
$v$ satisfying the linear inhomogeneous equation (\ref{inhomogeneous-eq})
obtained in \cite{PelSak}:
$$
v(t) = t \dot{\varphi}(t)+\frac{T(E)}{T'(E)} \partial_E \varphi(t;E),
$$
where $\varphi(t;E)$ denotes the family of even $T(E)$-periodic solutions
of the nonlinear oscillator equation (\ref{oscillator}) satisfying
the constraint $\varphi(0;E) = a(E) > 0$, where
$a(E)$ is the first positive root of $V(a(E)) = E$. Substituting $v$ into the formula for $A$,
we obtain
\begin{equation}
\label{A}
A = \frac{T(E)}{T'(E)} \left[ \ddot{\varphi}(t) \partial_E \varphi(t;E)
- \dot{\varphi} \partial_E \dot{\varphi}(t;E) \right] = - \frac{T(E)}{T'(E)},
\end{equation}
where the last identity holds because the time-independent Wronskian of
two linearly independent solutions of the homogeneous equation $\ddot{u} + V''(\varphi) u = 0$
is given explicitly by
\begin{eqnarray}
W &:=& \ddot{\varphi}(t) \partial_E \varphi(t;E)
- \dot{\varphi} \partial_E \dot{\varphi}(t;E) = \ddot{\varphi}(0) \partial_E \varphi(0;E)
- \dot{\varphi} \partial_E \dot{\varphi}(0;E)
\nonumber
\\
&=& \ddot{\varphi}(0) a'(E) = - V'(a(E)) a'(E) = -1.
\end{eqnarray}
Substituting (\ref{A}) into (\ref{Krein-first}), we obtain
\begin{equation}
\label{Krein-second}
K = -\frac{2 \Omega_0 T(E)}{T'(E)} \varepsilon^{\frac{1}{2}}
\sum_{j=1}^N |c_j|^2 + \mathcal{O}(\varepsilon).
\end{equation}
For the eigenvalue $\lambda = i \varepsilon^{\frac{1}{2}} \Omega_0 + \mathcal{O}(\varepsilon)$
with $\Omega_0$ given by  (\ref{internal-mode})
associated with the the internal mode of the two-site breather, we have $K > 0$ for the hard potential with $T'(E) < 0$
and $K < 0$ for the soft potential with $T'(E) < 0$.

Recall that $T(E) \to 2\pi$ as $E \to 0$, where $E \to 0$ represents the small-amplitude limit
in the nonlinear oscillator equation (\ref{oscillator}). We assume that $T'(E)$ remains sign-definite for all admissible
values of the energy parameter $E$. Let us now summarize on the Krein signatures for
the wave continuum and the internal mode of the two-site breathers:

\begin{itemize}
\item For the hard potential with $T'(E) < 0$, the internal mode with the Floquet multiplier
$\mu = e^{i T \left( \varepsilon^{\frac{1}{2}} \Omega_0 + \mathcal{O}(\varepsilon)\right)}$
has the positive Krein signature and so is the Krein signature for the wave spectrum
associated with the spectral band of
Floquet multipliers $\mu = e^{i T \left(1 + \mathcal{O}(\varepsilon)\right)}$. Since $T(E) < 2 \pi$
for the hard potentials, we have two situations:
\begin{itemize}
\item $0 < T < \pi$: the Krein signatures of the internal mode and the wave spectrum in the upper semi-circle coincide;
\item $\pi < T < 2\pi$: the Krein signatures of the internal mode and the wave spectrum in the upper semi-circle are opposite to each other.
\end{itemize}

\item For the soft potential with $T'(E) > 0$, the internal mode with the Floquet multiplier
$\mu = e^{i T \left( \varepsilon^{\frac{1}{2}} \Omega_0 + \mathcal{O}(\varepsilon)\right)}$
has the negative Krein signature, which is opposite to the Krein signature for the wave spectrum
corresponding to the spectral band of
Floquet multipliers $\mu = e^{i T \left(1 + \mathcal{O}(\varepsilon)\right)}$. Since $T(E) > 2 \pi$
for the soft potentials, we shall restrict out attention to the interval of $T$ between
$2 \pi$ and $4 \pi$ (which is the $1 : 2$ resonance).
If $2 \pi < T < 4 \pi$, we have two situations:
\begin{itemize}
\item $2\pi < T < 3\pi$: the Krein signatures of the internal mode and the wave spectrum in the upper semi-circle are opposite to each other;
\item $3\pi < T < 4 \pi$: the Krein signatures of the internal mode and the wave spectrum in the upper semi-circle coincide.
\end{itemize}
\end{itemize}
We will show that the nonlinear instability occurs when the Krein signatures of the internal mode and the wave spectrum are
opposite to each other. At the same time, the two-site breathers are
nonlinearly stable if the two signatures coincide.

\subsection{Asymptotic expansions}

We describe now the main result of this work. We show with the use of formal asymptotic
expansions that the multi-site breathers are unstable in the discrete KG equation (\ref{eq:dyn1d})
if the Krein signatures of the internal mode and the wave spectrum are opposite to each other.
To study the nonlinear dynamics of the internal mode, we adopt the asymptotic expansions
obtained earlier for the continuous and discrete NLS models \cite{KPS,pelkiv}.
We assume the following spectral properties:
\begin{itemize}
\item[P1] There exists a unique internal mode with eigenfrequency $\Omega$ such that
the Floquet multiplier $\mu = e^{i \Omega T}$ is isolated from the spectral bands located at
$\mu = e^{\pm i \omega(\theta) T}$, $\theta \in [-\pi,\pi]$, where $\omega(\theta)$
is given by (\ref{wave-spectrum}). To be mathematically precise, we assume that
\begin{equation}
\label{assumption-P1}
\Omega \in \left( 0, {\rm min}\{ 1 - k_0 \omega_0, m_0 \omega_0 - \sqrt{1 + 4 \varepsilon}\} \right),
\end{equation}
where $\omega_0 = \frac{2\pi}{T}$ is the breather frequency,
$k_0 \in \mathbb{N}$ is the maximal integer such that $1 - k_0 \omega_0 > 0$,
whereas $m_0 \in \mathbb{N}$ is the minimal integer such that $m_0 \omega_0 - \sqrt{1 + 4 \varepsilon} > 0$.

\item[P2] The double frequency of the internal mode eigenfrequency $\Omega$
belongs to the spectral band of the wave spectrum, that is,
\begin{equation}
\label{assumption-P2}
\mbox{\rm either} \;\; 2 \Omega \in (1- k_0 \omega_0,\sqrt{1 + 4 \varepsilon}- k_0 \omega_0) \quad \mbox{\rm or} \quad
2 \Omega \in (m_0 \omega_0 - \sqrt{1 + 4 \varepsilon},  m_0 \omega_0 - 1),
\end{equation}
where $\omega_0$, $k_0$, and $m_0$ are the same as for (\ref{assumption-P1}).

\item[P3] There exists a unique eigenvector $\dot{\bf u} \in C^{\infty}_{\rm per}((0,T);\ell^2(\mathbb{Z}))$
of the spectral problem (\ref{spectrumKG}) with $\lambda = 0$, where ${\bf u} \in C^{\infty}_{\rm per}((0,T);\ell^2(\mathbb{Z}))$
is the breather of the discrete KG equation (\ref{eq:dyn1d}).

\item[P4] The $T$-periodic functions $\dot{u}_n$ and $V'''(u_n) |W_n|^2$ are respectively odd and even
in the time variable $t$ for every $n \in \mathbb{Z}$, where ${\bf W} \in C^{\infty}_{\rm per}((0,T);\ell^2(\mathbb{Z}))$
is the internal mode for the eigenvalue $\lambda = i \Omega$ found from the spectral problem (\ref{spectrumKG}).
\end{itemize}

We introduce a small parameter $\delta$ that stands for the amplitude of
the internal mode. The small parameter $\delta$ is unrelated with
the possibly small parameter $\varepsilon$. Then, we look
for an asymptotic expansion
\begin{equation}
\label{delta-expansion}
{\bf U}(t) = {\bf u}(t) + \delta {\bf u}^{(1)}(t) + \delta^2 {\bf u}^{(2)}(t) + \delta^3 {\bf u}^{(3)}(t) + \cdots,
\end{equation}
where ${\bf u} \in C^{\infty}_{\rm per}((0,T);\ell^2(\mathbb{Z}))$ is the underlying breather.
We choose the first-order correction term in the form,
\begin{equation}
{\bf u}^{(1)}(t) = c(\tau) {\bf W}(t) e^{i \Omega t} + \bar{c}(\tau) \bar{\bf W}(t) e^{-i \Omega t},
\quad \tau = \delta^2 t,
\end{equation}
where $c$ is the slowly varying amplitude, ${\bf W} \in C^{\infty}_{\rm per}((0,T);\ell^2(\mathbb{Z}))$
is the eigenvector of the spectral problem (\ref{spectrumKG}) for the eigenvalue $\lambda = i \Omega$
isolated from the wave spectrum, see property (P1).

Separating the variables for the second-order correction term, we represent
\begin{equation}
{\bf u}^{(2)}(t) = c(\tau)^2 {\bf P}(t) e^{2 i \Omega t} + |c(\tau)|^2 {\bf Q}(t)
+ \bar{c}(\tau)^2 \bar{\bf P}(t) e^{-2 i \Omega t},
\end{equation}
where ${\bf P}, {\bf Q} \in C^{\infty}_{\rm per}((0,T);\ell^{\infty}(\mathbb{Z}))$
are solutions of the linear inhomogeneous equations
\begin{equation}
\label{linear-1}
\ddot P_n + 4 i \Omega \dot{P}_n - 4 \Omega^2 P_n + V''(u_n) P_n = \varepsilon(P_{n+1} - 2P_n + P_{n-1}) -\frac{1}{2} V'''(u_n) W_n^2,
\quad n \in \mathbb{Z}
\end{equation}
and
\begin{equation}
\label{linear-2}
\ddot Q_n + V''(u_n) Q_n = \varepsilon(Q_{n+1} - 2Q_n + Q_{n-1}) -  V'''(u_n) |W_n|^2,
\quad n \in \mathbb{Z}.
\end{equation}

Although the linear operator in the linear equation (\ref{linear-2}) is not invertible because
$\dot{\bf u} \in C^{\infty}_{\rm per}((0,T);\ell^2(\mathbb{Z}))$ is a solution of the homogeneous equation,
see property (P3), the Fredholm solvability condition is satisfied
$$
\sum_{n \in \mathbb{Z}}  \int_0^T V'''(u_n) |W_n|^2 \dot{u}_n dt = 0,
$$
because $\dot{u}_n$ is odd $T$-periodic and $V'''(u_n) |W_n|^2$ is even $T$-periodic functions with respect to the time variable $t$
for every $n \in \mathbb{Z}$, see property (P4). Consequently, there is a unique even $T$-periodic solution
${\bf Q}  \in C^{\infty}_{\rm per}((0,T);\ell^2(\mathbb{Z}))$ of the linear inhomogeneous equation (\ref{linear-2}).

The linear operator in the linear equation (\ref{linear-1}) is not invertible
because $2 \Omega$ belongs to the spectral band of the wave spectrum, see property (P2).
Consequently, the bounded even $T$-periodic solution ${\bf P} \in C^{\infty}_{\rm per}((0,T);\ell^{\infty}(\mathbb{Z}))$
is not uniquely defined unless the boundary conditions are added as $n \to \pm \infty$.
To obtain a unique bounded solution for ${\bf P}$, we specify the Sommerfeld radiation boundary conditions as follows.

Let $\sigma = 1$ if $T \in (0,\pi) {\rm mod}(2\pi)$ and $\sigma = -1$ if $T \in (\pi,2\pi) {\rm mod}(2\pi)$.
Then, we require that the even $T$-periodic solution ${\bf P} \in C^{\infty}_{\rm per}((0,T);\ell^{\infty}(\mathbb{Z}))$
of the linear inhomogeneous equation (\ref{linear-1}) satisfies the boundary conditions
\begin{equation}
\label{Sommerfeld}
R_{\pm}(t) = \lim_{n \to \pm \infty} P_n(t) e^{\pm i \sigma n \theta_0},
\end{equation}
where $R_{\pm} \in C^{\infty}_{\rm per}(0,T)$ are uniquely defined and
$\theta_0 \in (0,\pi)$ is uniquely found from the solution of the transcendental equations:
\begin{equation}
\label{theta-value}
\sigma = 1 : \quad  \omega(\theta_0) - k_0 \omega_0 = 2 \Omega \quad
\mbox{\rm or} \quad \sigma = -1 : \quad m_0 \omega_0 -  \omega(\theta_0) = 2 \Omega,
\end{equation}
where $\omega(\theta)$ is given by the dispersion relation (\ref{wave-spectrum}), $\omega_0 = \frac{2\pi}{T}$ is the breather frequency,
$k_0 \in \mathbb{N}$ is the maximal integer such that $1 - k_0 \omega_0 > 0$,
whereas $m_0 \in \mathbb{N}$ is the minimal integer such that $m_0 \omega_0 - \sqrt{1 + 4 \varepsilon} > 0$.
The existence of a unique $\theta_0 \in (0,\pi)$ follows from the assumption (P2).
We assume (and this will be a subject of the forthcoming work)
that there exists a unique even ${\bf P} \in C^{\infty}_{\rm per}((0,T);\ell^{\infty}(\mathbb{Z}))$
satisfying (\ref{linear-1}) and (\ref{Sommerfeld}).

The Sommerfeld radiation boundary conditions (\ref{Sommerfeld}) have the following physical meaning.
If $\sigma = 1$ and $\varepsilon$ is sufficiently small, the spectral band for the wave spectrum
corresponding to the Floquet multipliers $\mu = e^{i \omega(\theta) T}$, $\theta \in [-\pi,\pi]$ is located
in the upper semi-circle. Then, as $n \to \pm \infty$, the asymptotic solution (\ref{delta-expansion})
with the radiation boundary conditions (\ref{Sommerfeld}) becomes
$$
U_n(t) \sim \delta^2 \left( R_{\pm}(t) c^2(\tau) e^{\mp i n \theta_0 + i \left( \omega(\theta_0) - k_0 \omega_0 \right) t}
+ \bar{R}_{\pm}(t) \bar{c}^2(\tau) e^{\pm i n \theta_0 - i\left( \omega(\theta_0) - k_0 \omega_0 \right) t} \right) + \mathcal{O}(\delta^3).
$$
Recall that $\theta_0 \in (0,\pi)$ and $\omega'(\theta_0) > 0$, where the derivative of $\omega$ in $\theta$
determines the group velocity of the linear wave packets. Therefore, the linear wave packets
far from the breather field propagate outwards from $n = 0$, where
the breather is localized. On the other hand, if $\sigma = -1$, the spectral band for the wave spectrum
corresponding to the Floquet multipliers $\mu = e^{i \omega(\theta) T}$, $\theta \in [-\pi,\pi]$ is now located
in the lower semi-circle, so that the Floquet multiplier $\mu = e^{2 i \Omega}$ for the double frequency $2\Omega$
is in resonance with the spectral band corresponding to the Floquet multipliers $\mu = e^{-i \omega(\theta) T}$, $\theta \in [-\pi,\pi]$.
Then, as $n \to \pm \infty$, the asymptotic solution (\ref{delta-expansion})
with radiation boundary conditions (\ref{Sommerfeld}) becomes
$$
U_n(t) \sim \delta^2 \left( R_{\pm}(t) c^2(\tau) e^{\pm i n \theta_0 + i \left( m_0 \omega_0 -  \omega(\theta_0) \right) t}
+ \bar{R}_{\pm}(t) \bar{c}^2(\tau) e^{\mp i n \theta_0 - i \left( m_0 \omega_0 -  \omega(\theta_0) \right) t} \right) + \mathcal{O}(\delta^3).
$$
After the change in the sign of the Sommerfeld boundary conditions (\ref{Sommerfeld}),
the linear wave packets far from the breather field still propagate outwards from $n = 0$, where
the breather is localized.

Finally, separating the variables for the third-order correction term, we represent
\begin{equation}
{\bf u}^{(3)}(t) = c(\tau)^3 {\bf G}(t) e^{3 i \Omega t} + {\bf F}(t,\tau) e^{i \Omega t} +
\bar{\bf F}(t,\tau) e^{-i \Omega t} + \bar{c}(\tau)^3 \bar{\bf G}(t) e^{-3 i \Omega t},
\end{equation}
where ${\bf F}(\cdot,\tau), {\bf G} \in C^{\infty}_{\rm per}((0,T);\ell^{\infty}(\mathbb{Z}))$
are solutions of the linear inhomogeneous equations for every $\tau$.
We only write the problem for ${\bf F}$, because it yields a solvability condition
on the slowly varying amplitude $c$ of the internal mode. The function ${\bf F}$ is
found from the linear inhomogeneous equation:
\begin{equation}
\label{linear-3}
\ddot F_n + 2 i \Omega \dot{F}_n - \Omega^2 F_n + V''(u_n) F_n =
\varepsilon(F_{n+1} - 2F_n + F_{n-1}) + H_n,\quad n \in \mathbb{Z}
\end{equation}
where the dots denote derivatives with respect to $t$ and the source term ${\bf H}$ is given by
\begin{equation}
\label{linear-4}
H_n = - 2 \dot{c} \left( \dot{W}_n + i \Omega W_n \right) - |c|^2 c V'''(u_n)
(W_n Q_n + \bar{W}_n P_n) - \frac{1}{2} |c|^2 c V''''(u_n) |W_n|^2 W_n,
\quad n \in \mathbb{Z}.
\end{equation}
The linear operator in the linear equation (\ref{linear-3}) is not invertible because
${\bf W} \in C^{\infty}_{\rm per}((0,T);\ell^2(\mathbb{Z}))$ is a homogeneous solution,
see property (P1).
Therefore, a solution exists for ${\bf F}(\cdot,\tau) \in C^{\infty}_{\rm per}((0,T);\ell^2(\mathbb{Z}))$
if and only if the source term ${\bf H}$ satisfies the Fredholm solvability condition
$$
\frac{1}{T}  \sum_{n \in \mathbb{Z}}  \int_0^T \bar{W}_n(t) H_n(t,\tau) dt = 0,
$$
which yields the amplitude equation
\begin{equation}
\label{amplitude-eq}
i K \frac{dc}{d \tau} + \beta |c|^2 c = 0,
\end{equation}
where $K$ is the same as the time-independent Krein quantity (\ref{Krein-Omega}),
$$
K = -\frac{2i}{T} \sum_{n \in \mathbb{Z}}  \int_0^T \bar{W}_n \left( \dot{W}_n + i \Omega W_n \right) dt
= 2 \Omega \sum_{n \in \mathbb{Z}} |W_n|^2 + i \sum_{n \in \mathbb{Z}} (\dot{\bar{W}}_n W_n - \dot{W}_n \bar{W}_n),
$$
whereas $\beta$ is the coefficient of the cubic term given by
$$
\beta = \frac{1}{T}  \sum_{n \in \mathbb{Z}}  \int_0^T \left( V'''(u_n)
(|W_n|^2 Q_n + \bar{W}_n^2 P_n) + \frac{1}{2} V''''(u_n) |W_n|^4 \right) dt.
$$
The coefficient $\beta$ is complex because the bounded but non-decaying vector ${\bf P}$ is complex-valued,
according to the radiation boundary conditions (\ref{Sommerfeld}).
We are only interested in the imaginary part of $\beta$, which can be computed by
using the linear inhomogeneous equation (\ref{linear-1}) and integration by parts:
\begin{eqnarray*}
2 i {\rm Im}(\beta) & = & \frac{1}{T}  \sum_{n \in \mathbb{Z}}  \int_0^T V'''(u_n) \left(
\bar{W}_n^2 P_n - W_n^2 \bar{P}_n \right) dt \\
& = & \frac{2}{T}  \sum_{n \in \mathbb{Z}}  \int_0^T
\left[ \bar{P}_n \left( \ddot{P}_n + 4 i \Omega \dot{P}_n
- \varepsilon (\Delta P)_n \right) - P_n \left( \ddot{\bar{P}}_n - 4 i \Omega \dot{\bar{P}}_n
- \varepsilon (\Delta \bar{P})_n \right) \right] dt \\
& = & \frac{2 \varepsilon}{T}  \sum_{n \in \mathbb{Z}}  \int_0^T
\left[ P_n (\Delta \bar{P})_n -  \bar{P}_n (\Delta P)_n \right] dt,
\end{eqnarray*}
where we have used notation $(\Delta P)_n = P_{n+1} - 2 P_n + P_{n-1}$ for the discrete Laplacian operator.
Note that the integration by parts in $t$ yields a vanishing result because
${\bf P} \in C^{\infty}_{\rm per}((0,T);\ell^{\infty}(\mathbb{Z}))$.
However, because $P_n$ does not decay to zero as $|n| \to \infty$,
we have a nonzero result for ${\rm Im}(\beta)$. Indeed, we can write
$$
P_n (\Delta \bar{P})_n -  \bar{P}_n (\Delta P)_n = S_n - S_{n-1},
\quad S_n := P_n (\bar{P}_{n+1} - \bar{P}_n) - \bar{P}_n (P_{n+1} - P_n), \quad n \in \mathbb{Z},
$$
interchange the integration and summation, and use the telescopic summations to obtain
\begin{eqnarray*}
2 i {\rm Im}(\beta) & = & \frac{2 \varepsilon}{T}  \int_0^T
\left[ \lim_{n \to +\infty} S_n - \lim_{n \to -\infty} S_n \right] dt.
\end{eqnarray*}
Substituting the Sommerfeld boundary conditions (\ref{Sommerfeld}), we arrive to the sign-definite
expression,
\begin{eqnarray*}
2 i {\rm Im}(\beta) & = &  4 \varepsilon \sigma \sin(\theta_0) \frac{1}{T} \int_0^T \left( |R_+(t)|^2 + |R_-(t)|^2 \right) dt,
\end{eqnarray*}
where $\sin(\theta_0) > 0$ for $\theta_0 \in (0,\pi)$. It follows that ${\rm Im}(\beta) \neq 0$,
if the radiation amplitudes $R_{\pm}$ in the boundary conditions (\ref{Sommerfeld}) are nonzero.
Multiplying (\ref{amplitude-eq}) by $\bar{c}$
and subtracting the complex conjugate equation, we obtain the rate of change for the squared amplitude:
\begin{equation}
\label{energy-eq}
K \frac{d |c|^2}{d \tau} = - 4 \varepsilon \sigma \sin(\theta_0) |c|^4 \frac{1}{T} \int_0^T \left( |R_+(t)|^2 + |R_-(t)|^2 \right) dt.
\end{equation}
It follows from this equation that the squared amplitude $|c|^2$ decays to zero
if ${\rm sign}(K) = {\rm sign}(\sigma)$ and grows if ${\rm sign}(K) = -{\rm sign}(\sigma)$.
This is the main result of the asymptotic theory. In the former case, we can anticipate that
the multi-site breathers are stable in the nonlinear dynamics of the discrete KG equation.
In the latter case, we can predict that the multi-site breathers are nonlinearly unstable, in spite
of their linearized stability.

%Let us recall the different cases from the end of Section II and summarize
%the stability predictions for the two-site breathers with the period $T$
%near the anti-continuum limit with small $\varepsilon > 0$:
%\begin{itemize}
%\item Nonlinear stability holds for the hard potential when $0 < T < \pi$
%and for the soft potential when $3\pi < T < 4 \pi$;
%\item Nonlinear instability holds for the hard potential when $\pi < T < 2 \pi$
%and for the soft potential when $2 \pi < T < 3 \pi$.
%\end{itemize}

In the approximation of the small-amplitude breathers with the discrete NLS equations \cite{PPP},
$T$ is close to $2\pi$, so both hard and soft potentials feature nonlinear instability of
the two-site breathers. This phenomenon was discovered recently for the discrete NLS
equation in \cite{KPS}, following upon the earlier abstract analysis
of the continuous NLS equation in \cite{Cuccagna}.
Note that the conclusion changes drastically when the breather
period is either smaller than $\pi$ or bigger than $3\pi$.
The former case is observed for small-period breathers in the hard potentials.
The latter case is observed for breathers in the soft potentials near the $1 : 2$ resonance.
In either case, the multi-site breathers are expected to be nonlinearly stable.

\section{Numerical results}

Having explored the fundamentals of the nonlinear instability of multi-site breathers, we now
turn to a numerical examination of the discrete KG equation (\ref{eq:dyn1d}) for two potentials
in (\ref{potentials}), namely the soft (Morse) and the hard ($\phi^4$) potentials.
We choose two-site breathers with the excited sites at $n=0$ and $n=1$
oscillating in anti-phase for soft potentials, so that $u_0(0)=-u_1(0)$,
and oscillating in phase for hard potentials, so that $u_0(0)=u_1(0)$. According to section 2.3,
these two-site breathers possess an internal mode with the small eigenfrequency
$\Omega = \varepsilon^{\frac{1}{2}} \Omega_0 + \mathcal{O}(\varepsilon)$,
where $\Omega_0$ is given by (\ref{internal-mode}). The corresponding pair of
Floquet multipliers $\mu = e^{\pm i \Omega T}$ is located near $\mu = 1$ for small $\varepsilon$.
The Krein signature, ${\rm sign}(K)$, of the internal mode for the Floquet multiplier
$\mu = e^{i \Omega}$ in the
upper half-circle is $+1$ if the potential $V$ is soft and $-1$ if $V$ is hard.

\begin{figure}[h]
\begin{center}
\begin{tabular}{cc}
Morse potential: $\omega_0 = 0.85$ & Morse potential: $\omega_0 = 0.65$ \\
\includegraphics[width=5.cm]{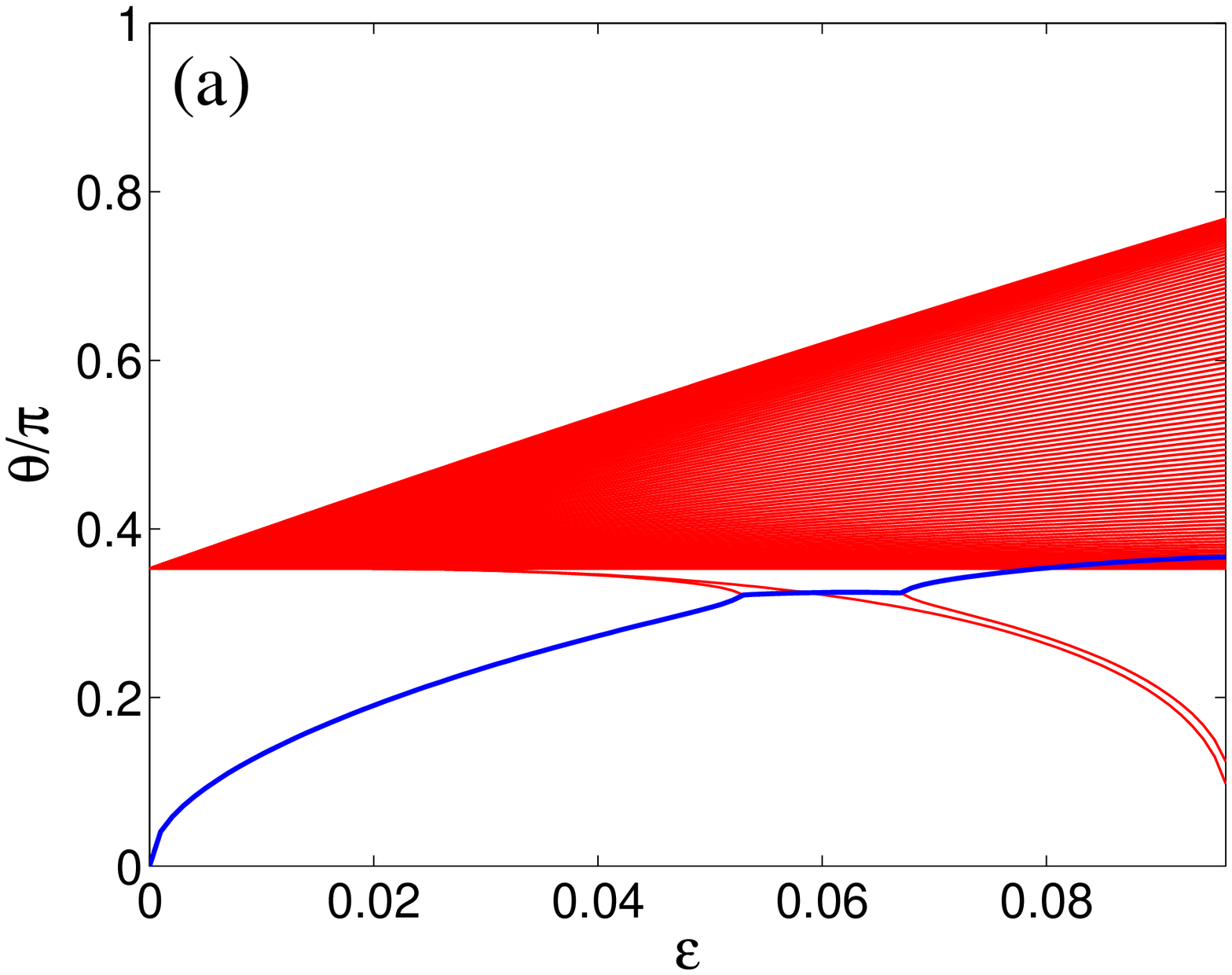} &
\includegraphics[width=5.cm]{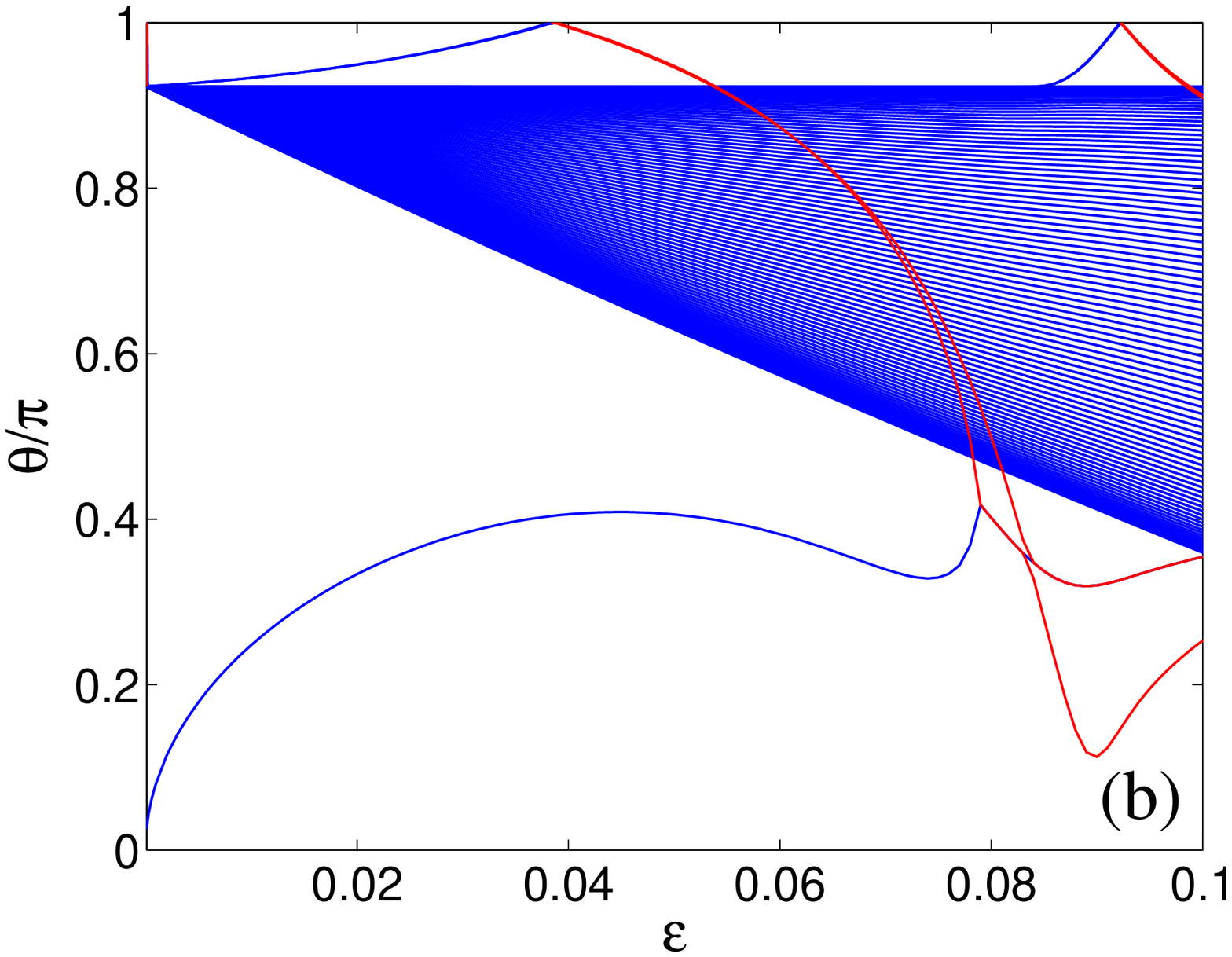} \\
$\phi^4$ potential: $\omega = 2$ & $\phi^4$ potential: $\omega_0 = 5$ \\
\includegraphics[width=5.cm]{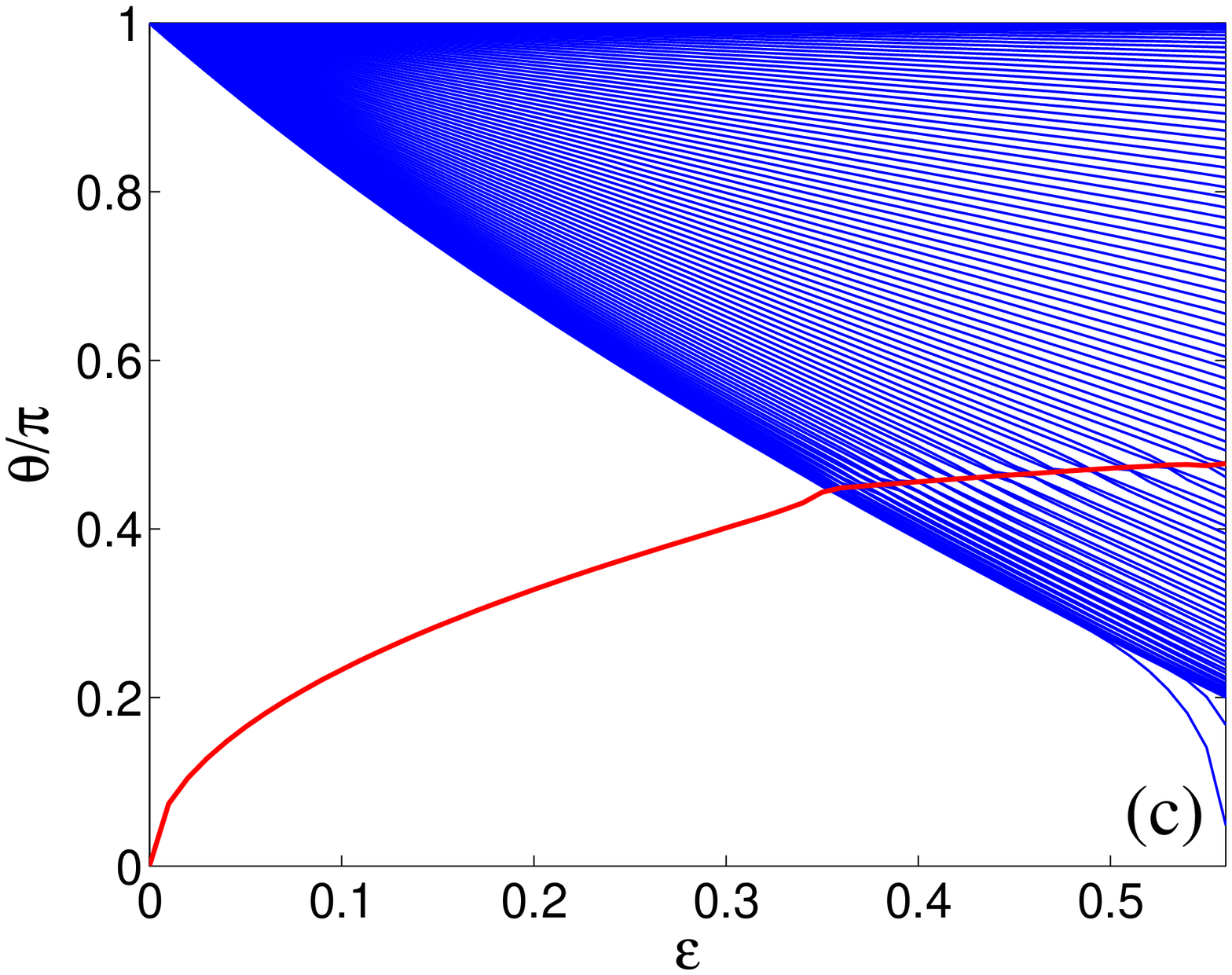} &
\includegraphics[width=5.cm]{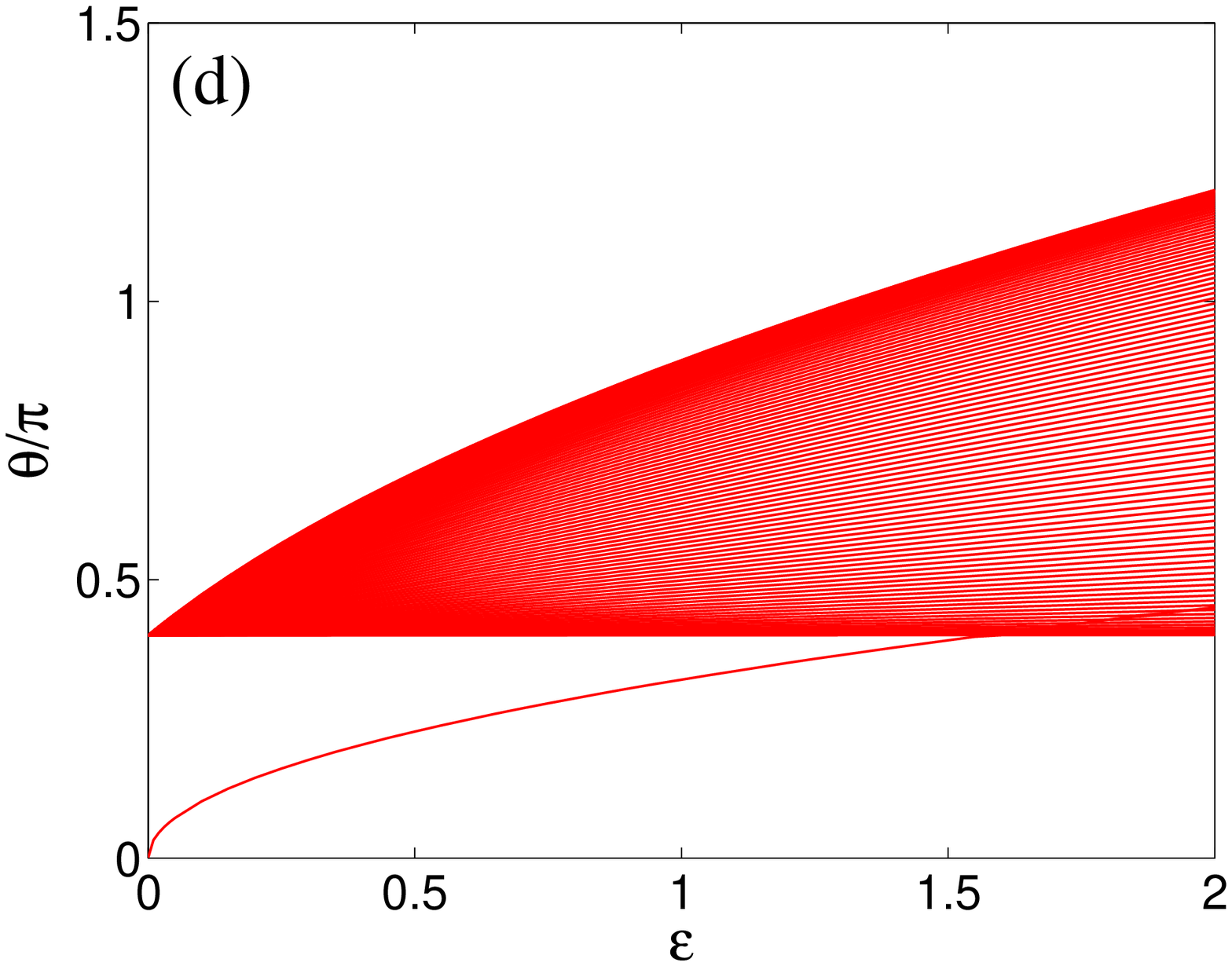} \\
\end{tabular}
\end{center}
\caption{Floquet argument $\theta$ restricted on $[0,\pi]$ versus $\varepsilon$
for the two-site breathers oscillating in anti-phase for the Morse potential (a,b) and
in phase for the $\phi^4$ potential (c,d). Red (blue) lines represent spectral modes
with positive (negative) Krein signature.}
\label{fig:angles}
\end{figure}

In general, independently of the expression for the potential $V$,
the pair of phonon arcs on the unit circle (which represent Floquet multipliers
for the wave continuum bands) does not overlap for small $\varepsilon$;
one arc is located in the upper half-circle and the other arc is located
in the lower half-circle. The Krein signatures
of the two distinct phonon arcs are opposite to each other.
The arc in the upper half-circle has ${\rm sign}(K) =-1$ if $\pi < T < 2\pi$ or
$3 \pi < T < 4 \pi$ and ${\rm sign}(K) =1$ if $0 < T < \pi$ or
$2 \pi < T < 3 \pi$, where $T$ is the breather period.
Recall that $T < 2 \pi$ for hard potentials and $T > 2 \pi$ for soft potentials.
For an earlier discussion of these spectral features, see e.g. \cite{MS98,MAF98}.

If $2\Omega$ lies inside the wave continuum, the breathers are subject to
the nonlinear instability mechanism presented herein,
as long as the Krein signature of the phonon arc
at the upper half-circle is the opposite of the Krein signature of the
internal mode residing in this half-circle. This implies that for the configurations above,
nonlinear instability can arise for the intervals
$\pi < T < 2 \pi$ (hard potentials) and $2 \pi < T < 3 \pi$ (soft potentials). However, nonlinear instability
cannot be  observed in the intervals  $0 < T < \pi$ (hard potentials) and $3 \pi < T < 4 \pi$ (soft potentials).
All four cases are represented in Fig. \ref{fig:angles}, where the dependence of
the relevant arguments $\theta$ of the Floquet multipliers $\mu = e^{i \theta}$
is displayed versus the coupling constant $\varepsilon$.
For further experiments, we represent the unstable and stable cases as follows:
\begin{itemize}
\item Morse potential: unstable case - $\varepsilon = 0.03$ for $\omega_0 = 0.85$ ($T \approx 2.35 \pi$);
stable case - $\varepsilon = 0.04$ for $\omega_0 = 0.65$ ($T \approx 3.08 \pi$).
\item $\phi^4$ potential: unstable case - $\varepsilon = 0.3$ for $\omega_0 = 2$ ($T = \pi$);
stable case -  $\varepsilon = 1$ for $\omega_0 = 5$ ($T = 0.4 \pi$).
\end{itemize}

In order to observe the emergence of the nonlinear instability, breathers
are perturbed by adding the internal mode
$({\bf{W}}(0),\dot{\bf{W}}(0))$, multiplied by a
relatively small factor $\delta$, to the breather solution
$({\bf{u}}(0),\dot{\bf{u}}(0))$.
The reported numerical integration results have been obtained
by means of the 4th order explicit and symplectic Runge-Kutta-Nystr\"om
method developed in~\cite{Calvo-paper,Calvo}. This scheme
preserves the energy up to a $\mathcal{O}(10^{-8})$ factor even for the long
integration times ($\gtrsim 10^4$) used here.

\begin{figure}
\begin{center}
\begin{tabular}{cc}
\includegraphics[width=5.cm]{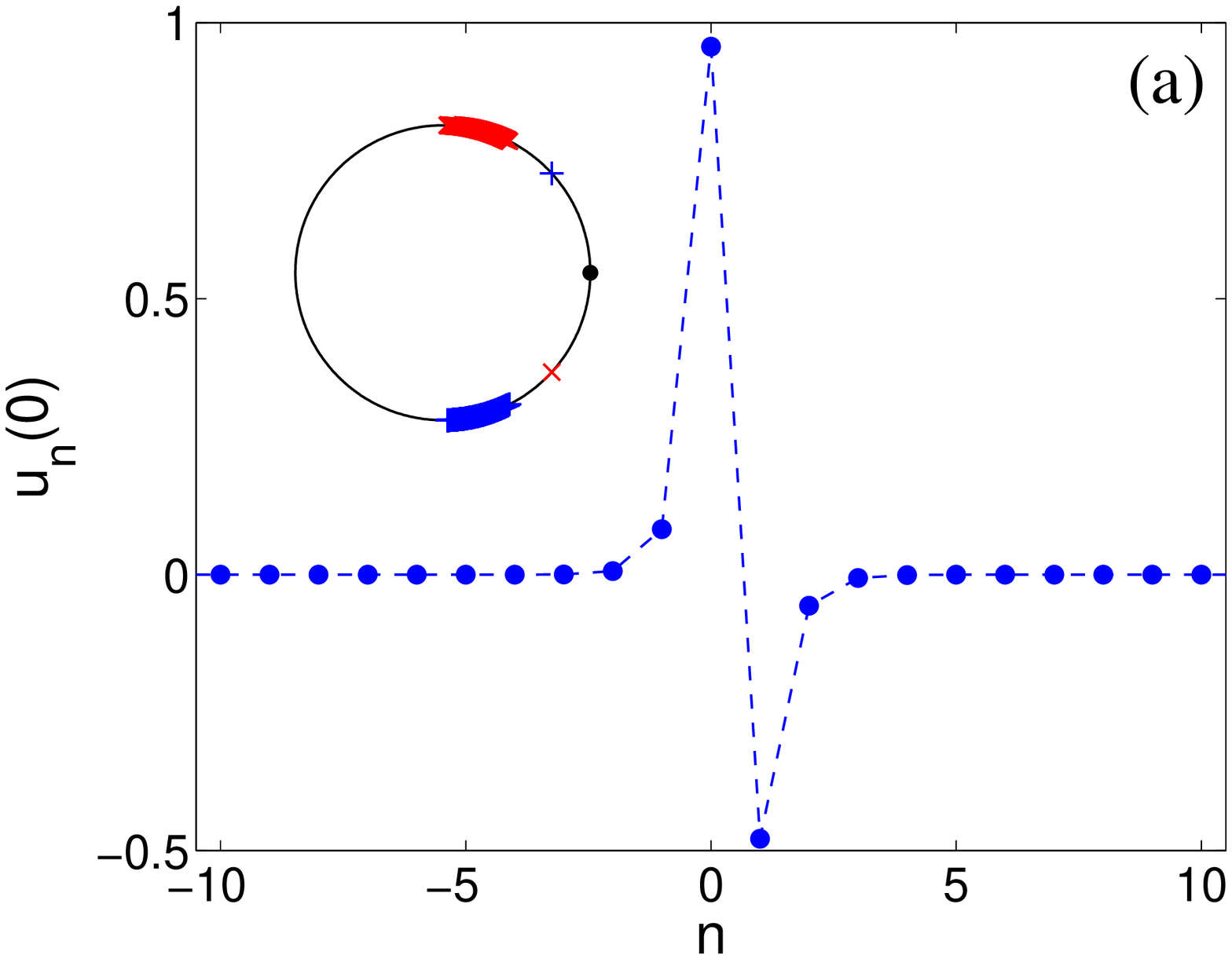} &
\includegraphics[width=5.cm]{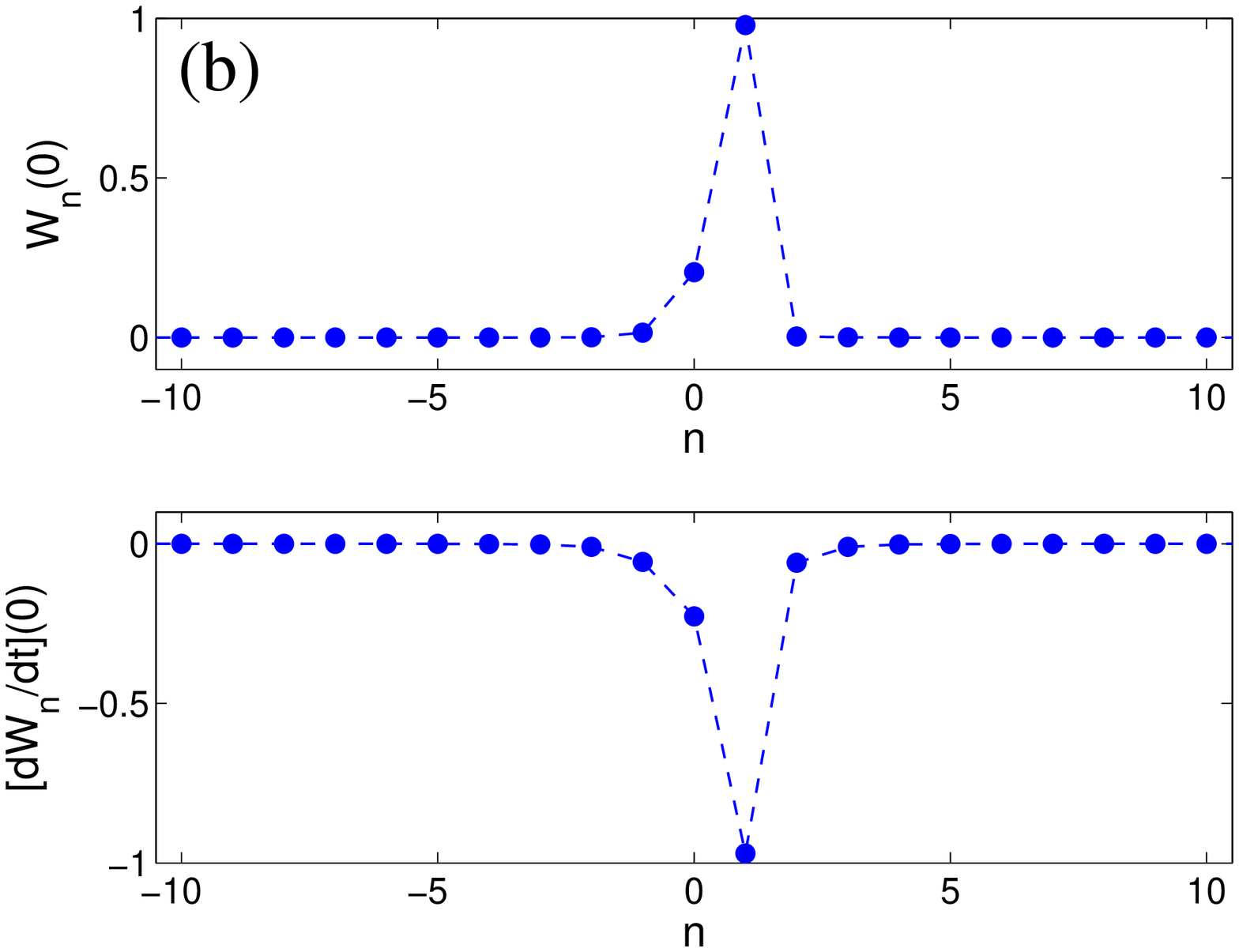} \\
\includegraphics[width=5.cm]{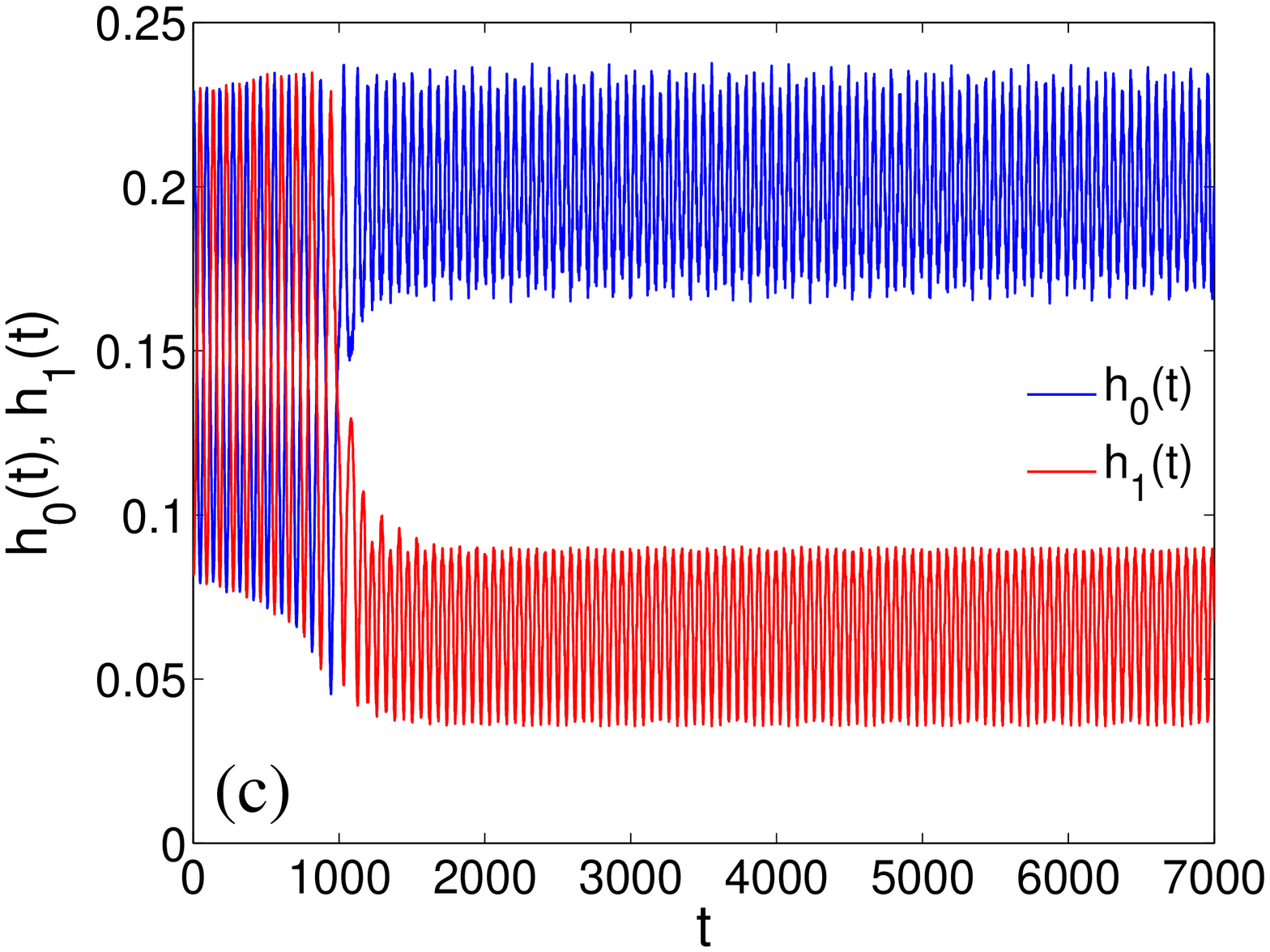} &
\includegraphics[width=5.cm]{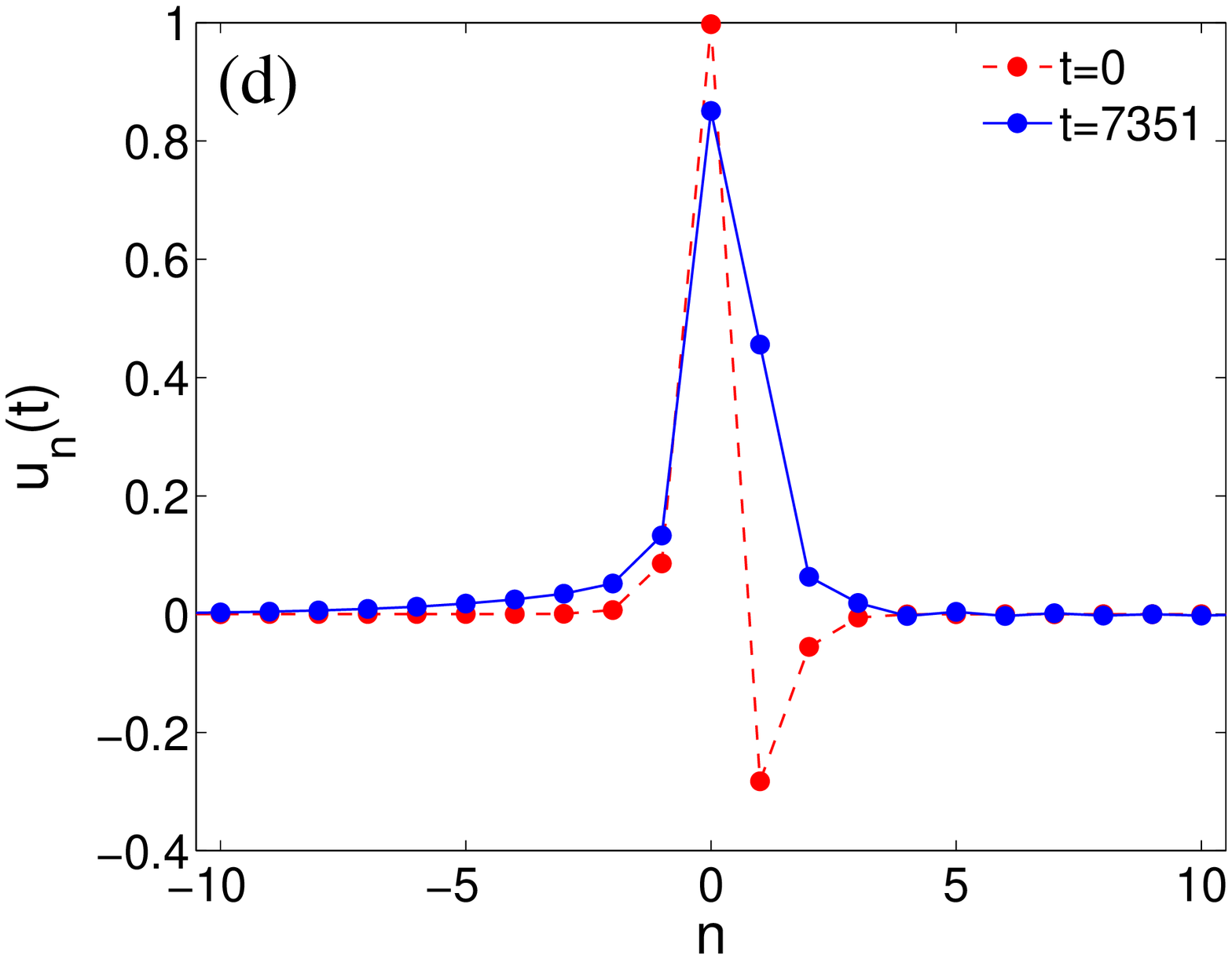} \\
\end{tabular}
\end{center}
\caption{The two-site anti-phase breather, predicted to be nonlinearly
unstable, in the Morse potential for $\varepsilon = 0.03$ for $\omega_0 = 0.85$:
the stationary breather profile and Floquet spectrum (a), the internal mode (b), the
energy density versus time (c), and the profile of the perturbed breather at $t=0$
and at the end of the simulation (d).}
\label{fig:Morse1}
\end{figure}

Fig. \ref{fig:Morse1} shows the outcome for dynamics of the two-site anti-phase breather in the
Morse potential for $\varepsilon = 0.03$ for $\omega_0 = 0.85$ ($T \approx 2.35 \pi$).
In this case, the internal mode eigenfrequency is $\Omega \approx 0.1002$, whereas the
nearest phonon band shifted by $k_0=1$ [see (\ref{assumption-P1}) and (\ref{assumption-P2})]
is located in the frequency range $[0.15,0.2083]$. As a result, the double frequency
$2 \Omega$ is inside the phonon band. The two-site breather is perturbed by the internal mode multiplied by $\delta=0.2$.

Panels (a) and (b) of Fig. \ref{fig:Morse1} include the breather profile with the Floquet spectrum
and the internal mode $({\bf{W}},\dot{\bf{W}})$ at $t = 0$. Panel (c) shows
the energy density at the central sites $n = 0,\ 1$. The energy density is defined by
\begin{equation}
    h_n=\frac{\dot u_n^2}{2}+V(u_n)+\frac{\varepsilon}{4}\left[(u_n-u_{n+1})^2+(u_n-u_{n-1})^2\right],
\end{equation}
so that the Hamiltonian in (\ref{eq:ham1d}) can be written as $H=\sum_{n \in \mathbb{Z}}h_n$.
%The evolution of the displacements of the central sites are displayed in panel (d), whereas
Panel (d) compares the profile of the perturbed breather at $t=0$ to the profile of the
breather near the end of the simulation.
%Finally, panel (f) shows the Fourier spectrum of $u_0$,
%denoted as $\hat u_0$ and calculated from the last $\sim750$ oscillations of the amplitude $u_0$.

From Fig. \ref{fig:Morse1}c, we observe the nonlinear instability of the two-site breather
with frequency $\omega_0 = 0.85$, which manifests by a sudden decay of the energy density
at the $n=1$ site and the growth of the energy density at the $n=0$ site. A
quasi-periodic single-site breather is formed as a result of this instability of the two-site breather.

%From Fig. \ref{fig:Morse1}f, we can observe the emergence of several new frequencies. In particular, the breather frequency becomes $\omega'_0=0.8463$. We also observe appearance of an additional modulation frequency $\omega_\mathrm{I}=0.1069$ that gives peaks at $\omega'_0\pm\omega_\mathrm{I}$. The new frequency of the two-site breather $\omega'_0$ is close to the initial frequency $\omega_0$.

\begin{figure}
\begin{center}
\begin{tabular}{cc}
\includegraphics[width=5.cm]{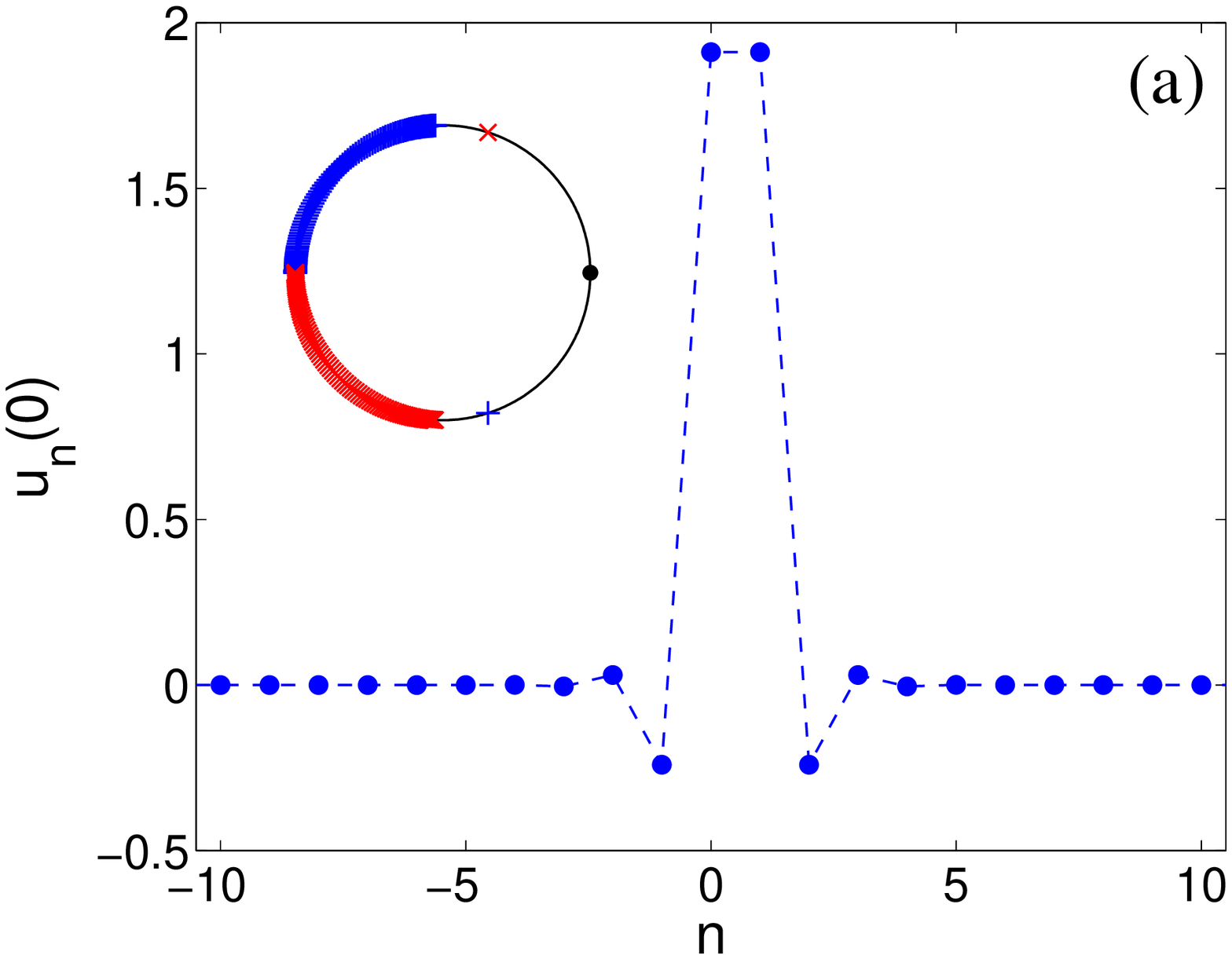} &
\includegraphics[width=5.cm]{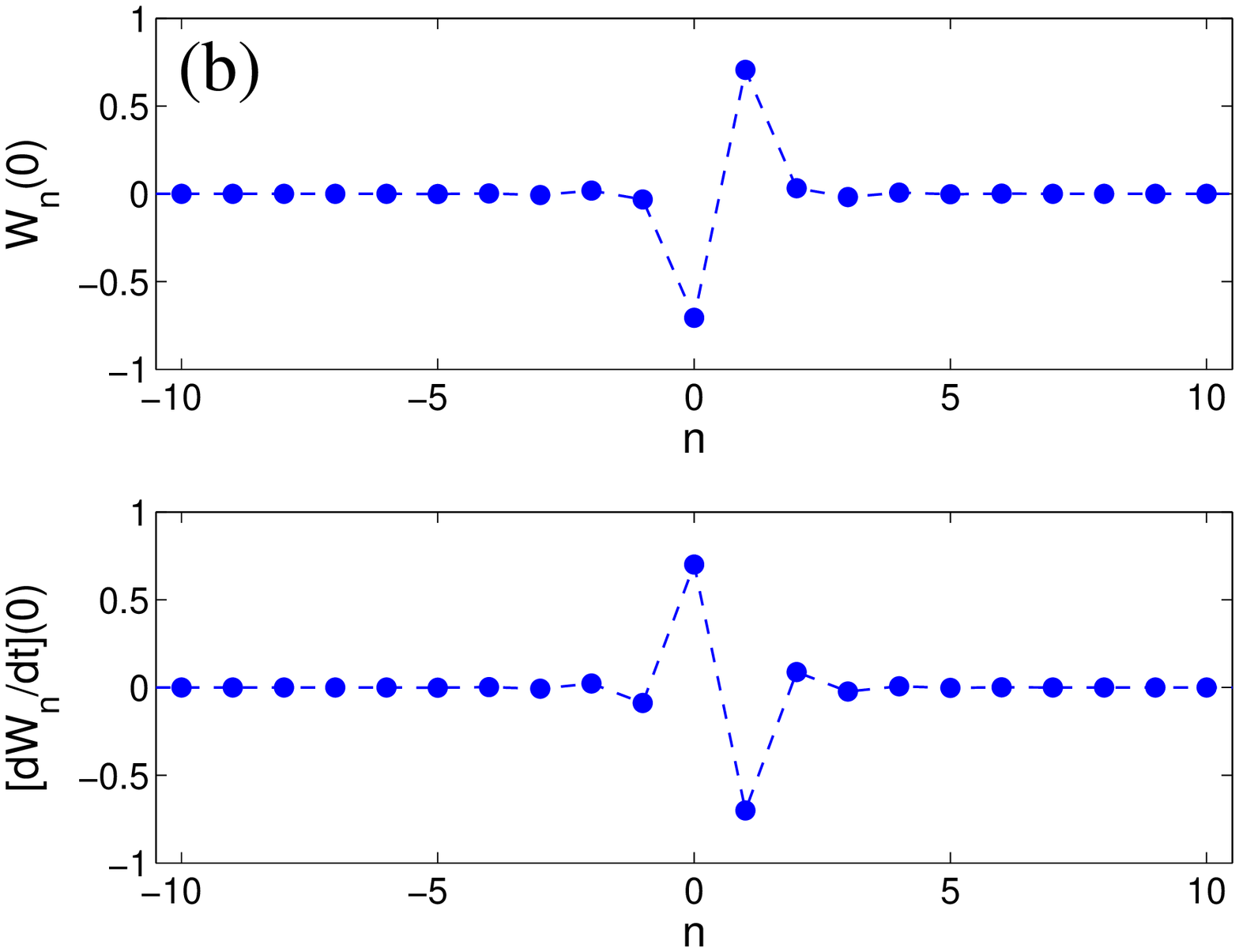} \\
\includegraphics[width=5.cm]{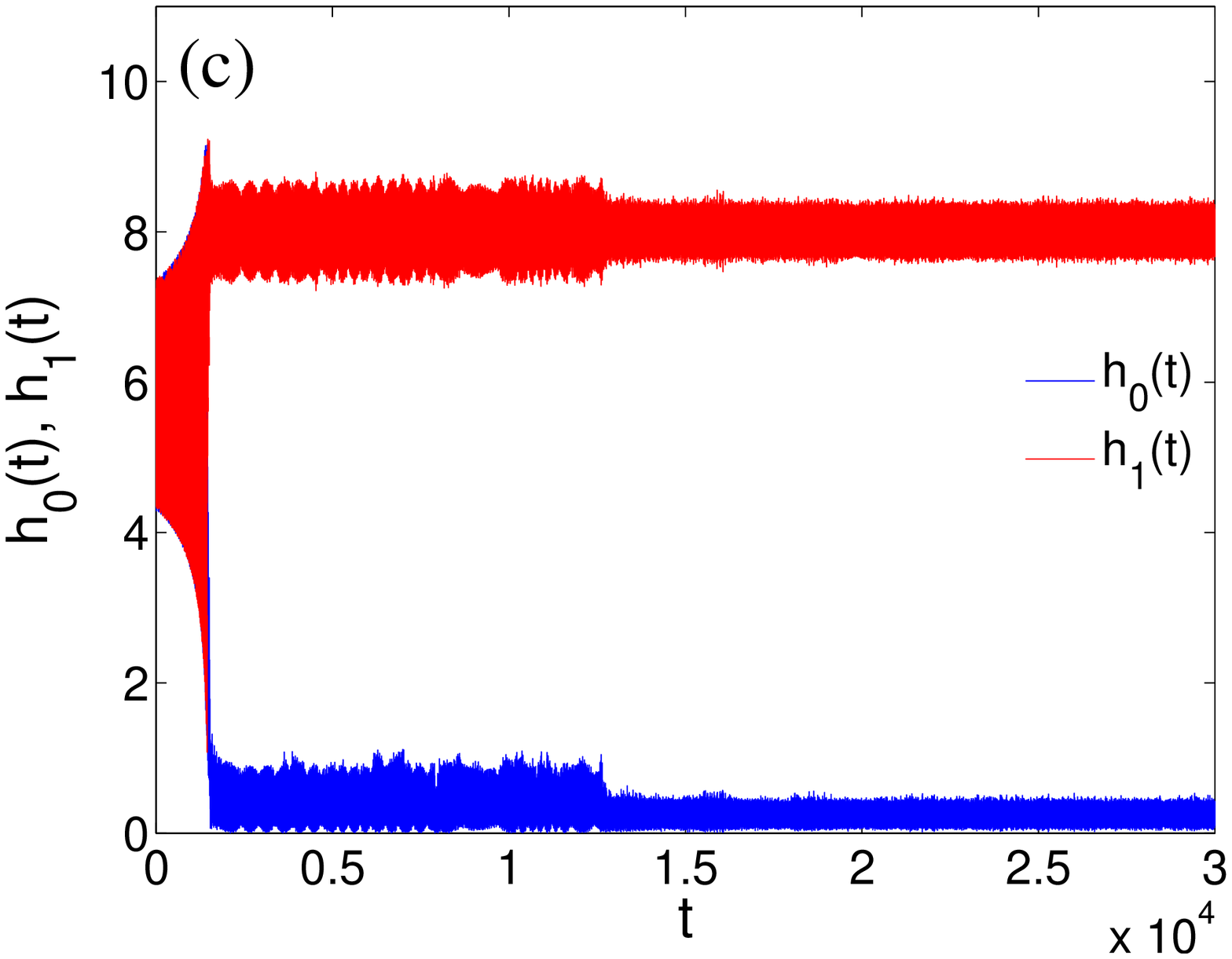} &
\includegraphics[width=5.cm]{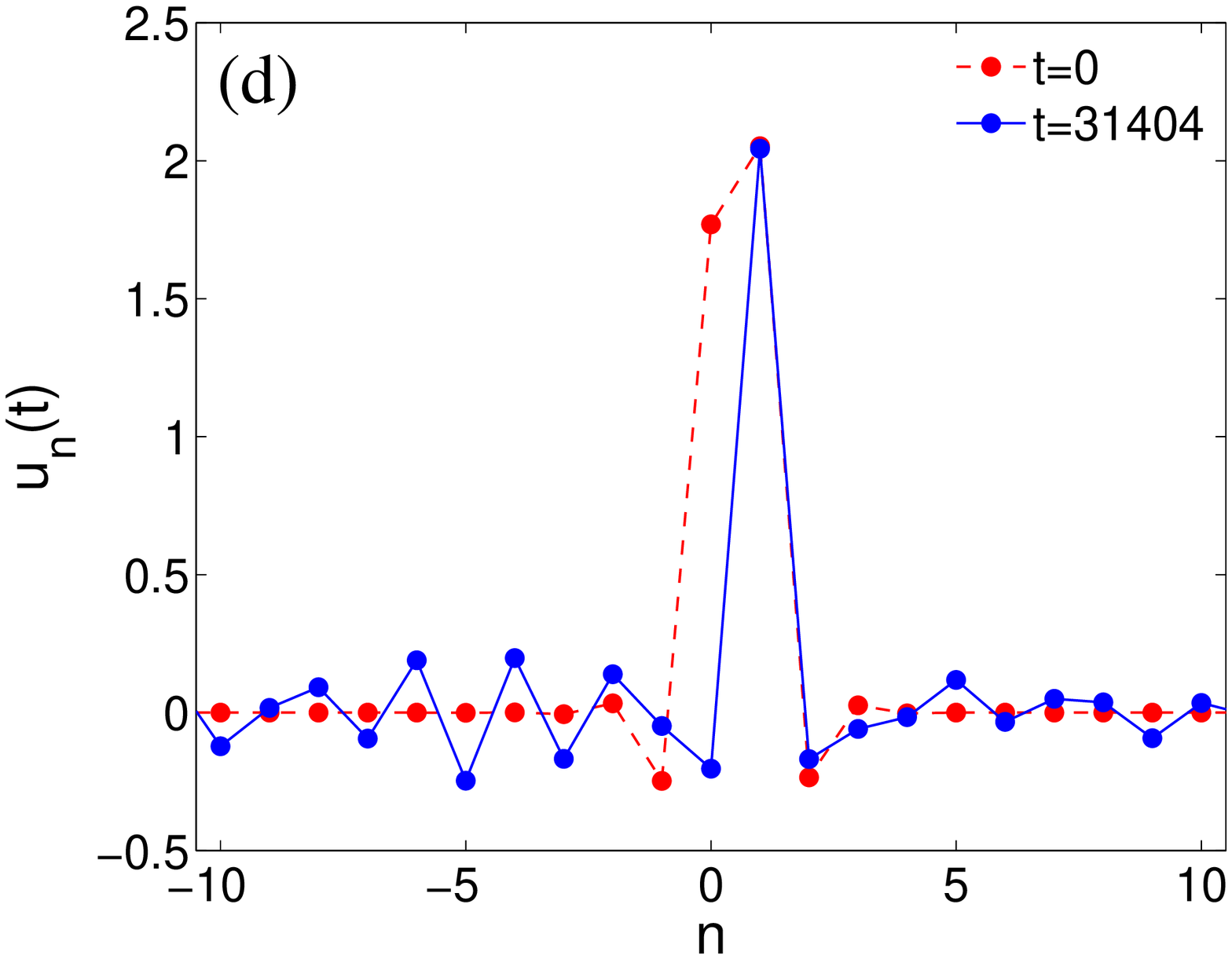} \\
\end{tabular}
\end{center}
\caption{Similar to Fig. \ref{fig:Morse1} but now for the two-site in-phase
breather in the hard $\phi^4$ potential for $\varepsilon=0.3$ and $\omega_0 = 2$.}
\label{fig:quartic1}
\end{figure}

Fig. \ref{fig:quartic1} shows the outcome for dynamics of the two-site in-phase breather in the
hard $\phi^4$ potential for $\varepsilon=0.3$, $\delta=0.2$, and $\omega_0 = 2$ ($T=\pi$).
%We have selected the marginal case $T = \pi$ compared to the more general case $T \in (\pi,2\pi)$ because
%the peaks of the Fourier spectrum are better defined and their number is smaller in the case $T = \pi$.
For this case, the internal mode eigenfrequency is $\Omega \approx 0.4011$, whereas the
nearest phonon band shifted by $m_0=1$ [see (\ref{assumption-P1}) and (\ref{assumption-P2})]
is located in the frequency range $[0.56,2]$. Again, the double frequency
$2 \Omega$ is inside the phonon band.
As a result, similarly to Fig. \ref{fig:Morse1}, we observe
the nonlinear instability of the two-site breather and its transformation to a quasi-periodic
single-site breather.
%Panel (f) shows that the new frequency of the single-site breather is $\omega'_0=2.1978$. It also shows a modulation frequency $\omega_\mathrm{I}=4.3952$.

We point out that the final state in both computations of Figs. \ref{fig:Morse1} and \ref{fig:quartic1}
consists of quasi-periodic oscillations of a single-site breather. This behavior appears to be
generic for the range $[\pi,2\pi]$ of the periods $T$ supported by the hard $\phi^4$ potential
and the range $[2 \pi, 3 \pi]$ of the periods $T$ supported by the soft Morse potential.

\begin{figure}
\begin{center}
\begin{tabular}{cc}
\includegraphics[width=5.cm]{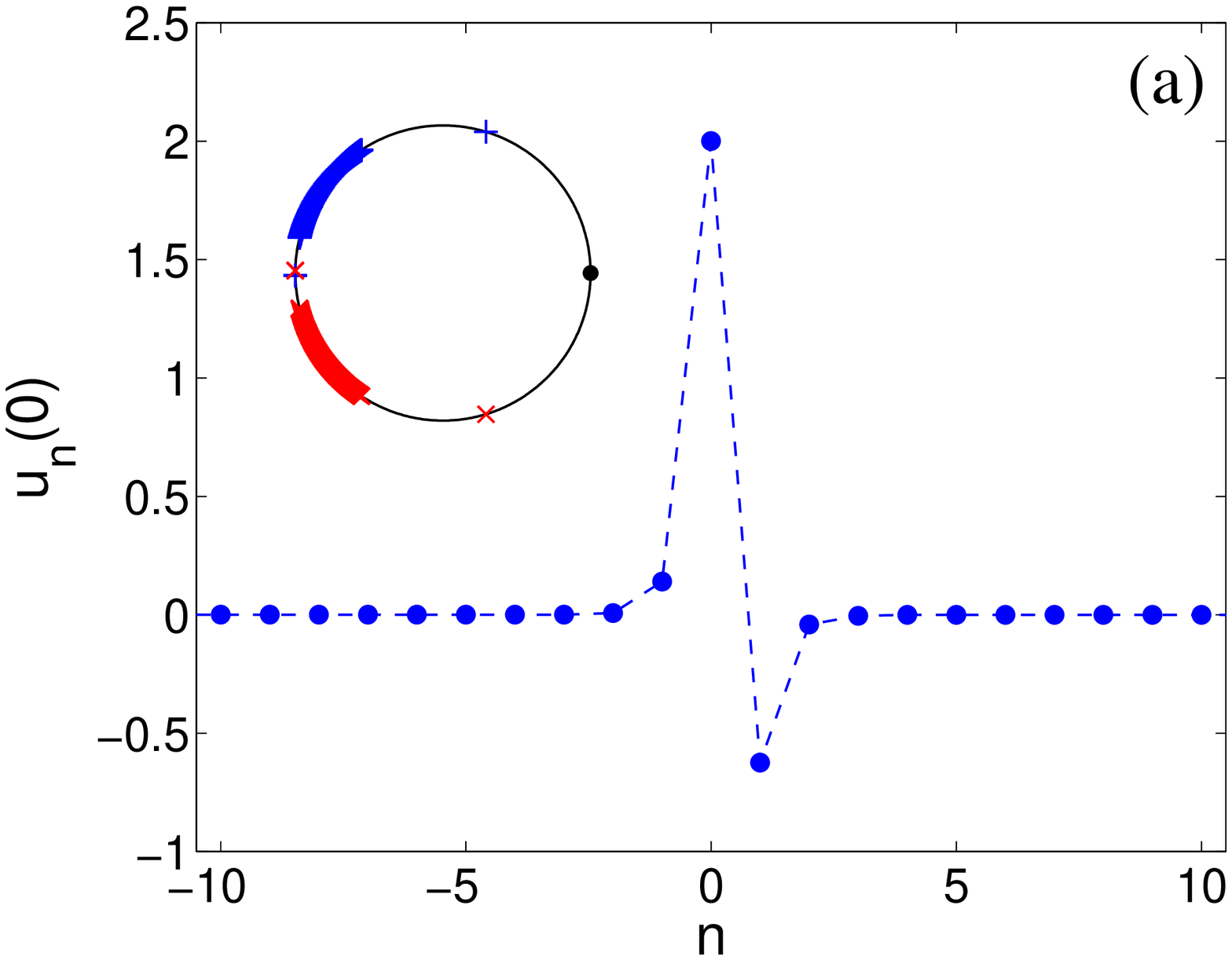} &
\includegraphics[width=5.cm]{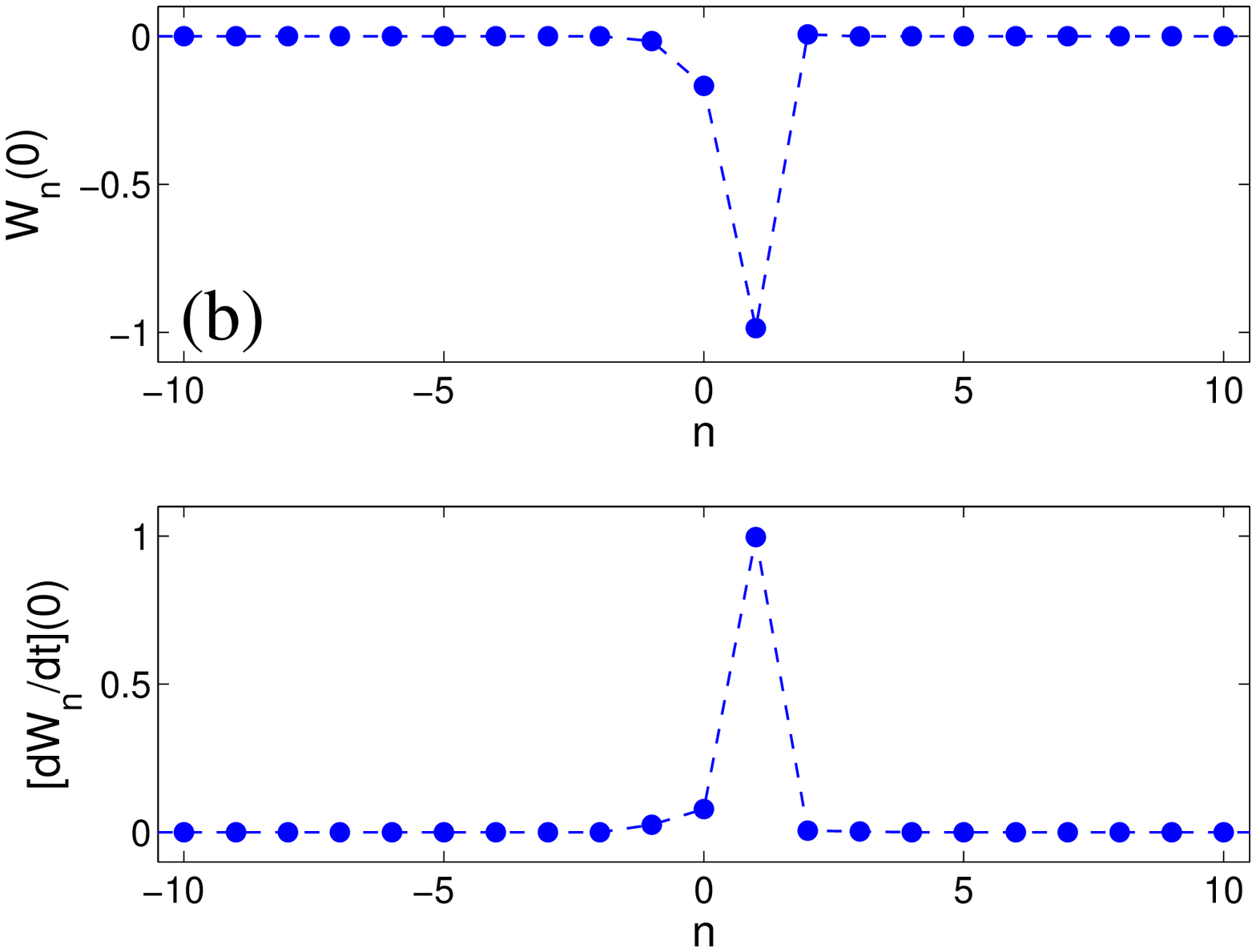} \\
\includegraphics[width=5.cm]{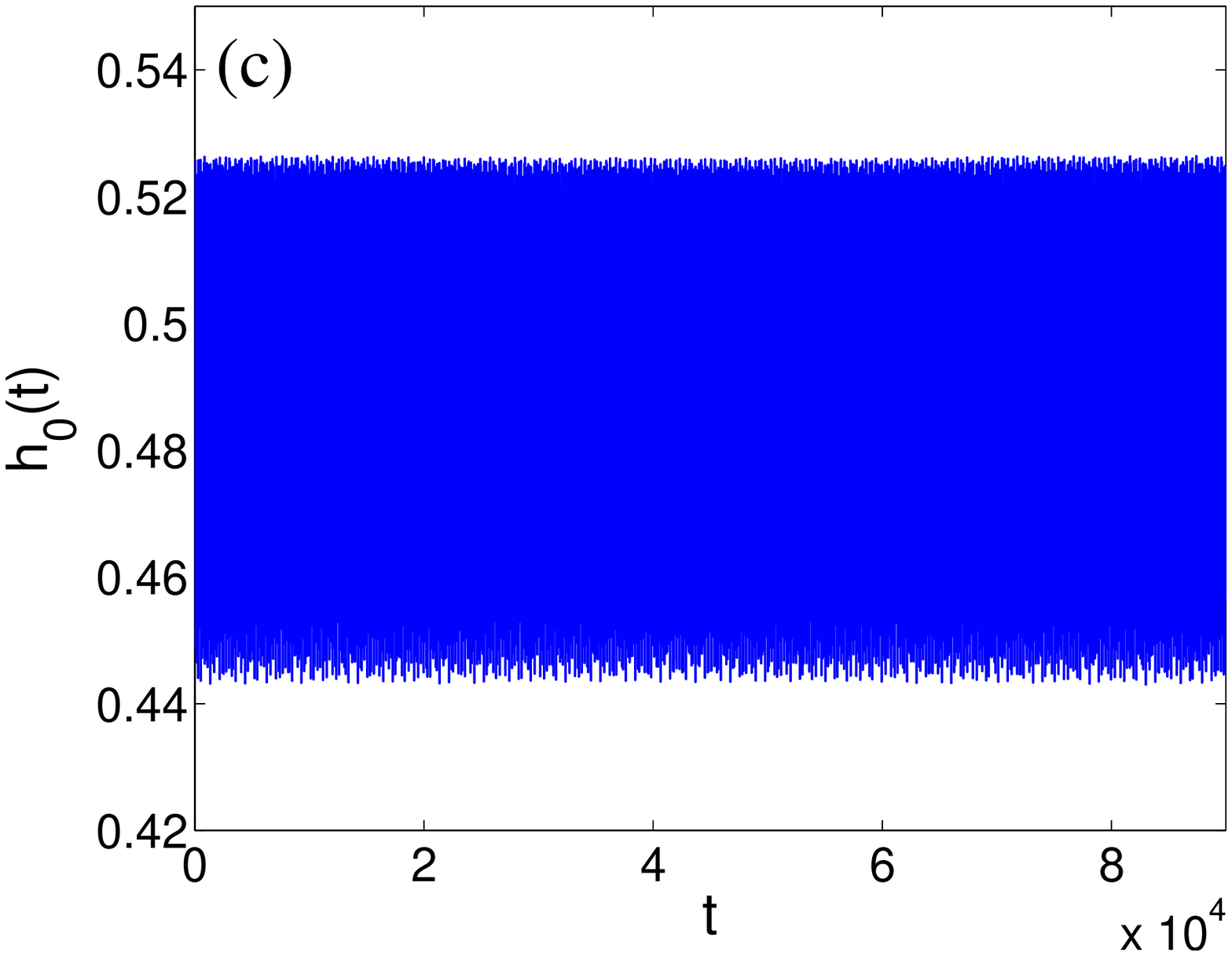} &
\includegraphics[width=5.cm]{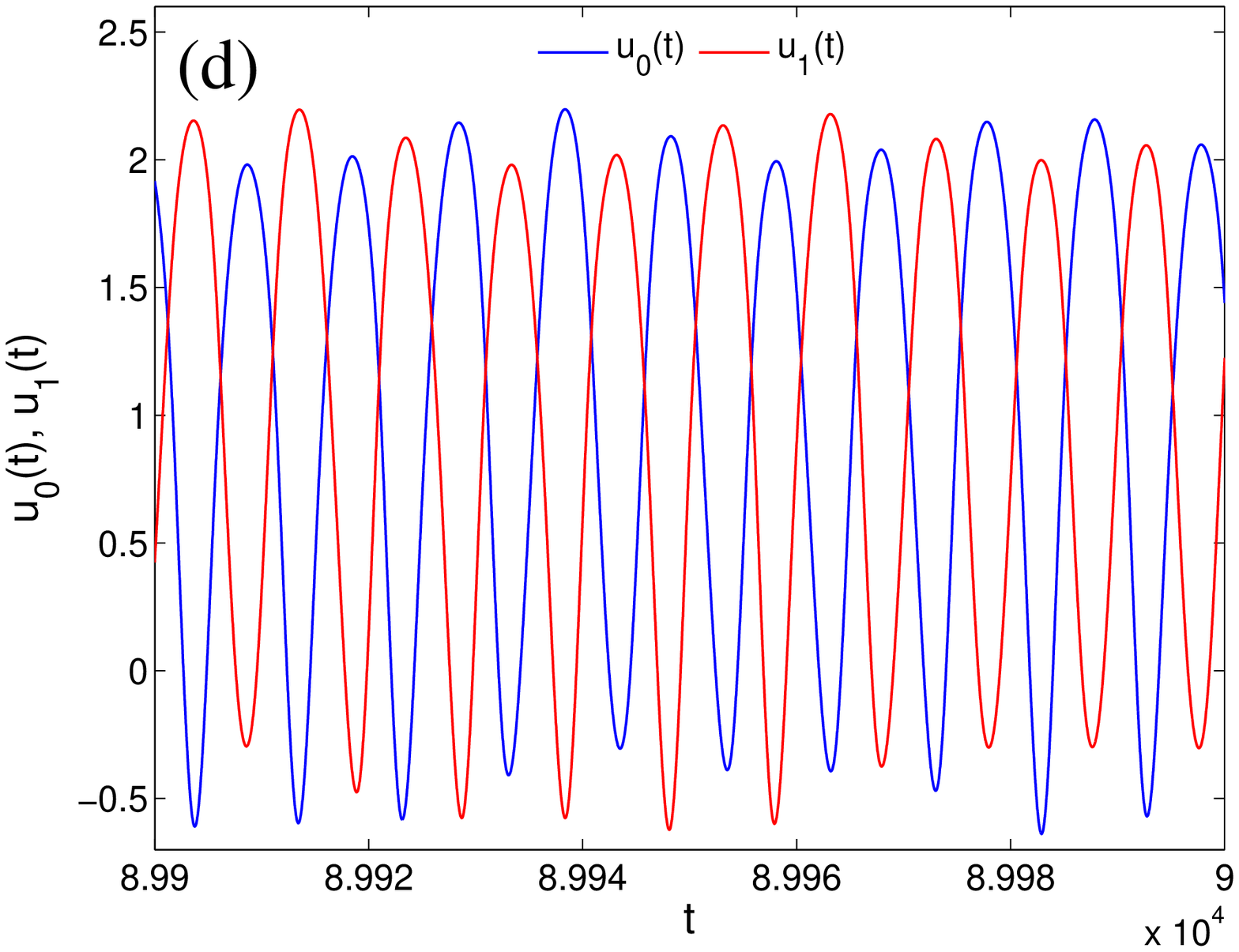} \\
\end{tabular}
\end{center}
\caption{The two-site anti-phase breather, predicted to be nonlinearly
stable, in the Morse potential for $\varepsilon = 0.04$ for $\omega_0 = 0.65$:
the stationary breather profile and Floquet spectrum (a), the internal mode
(b), and
the energy density of the central site versus time (c). Panel (d) shows the evolution
of the central sites $n=0$ and $n=1$.}
\label{fig:Morse2}
\end{figure}

\begin{figure}
\begin{center}
\begin{tabular}{cc}
\includegraphics[width=5.cm]{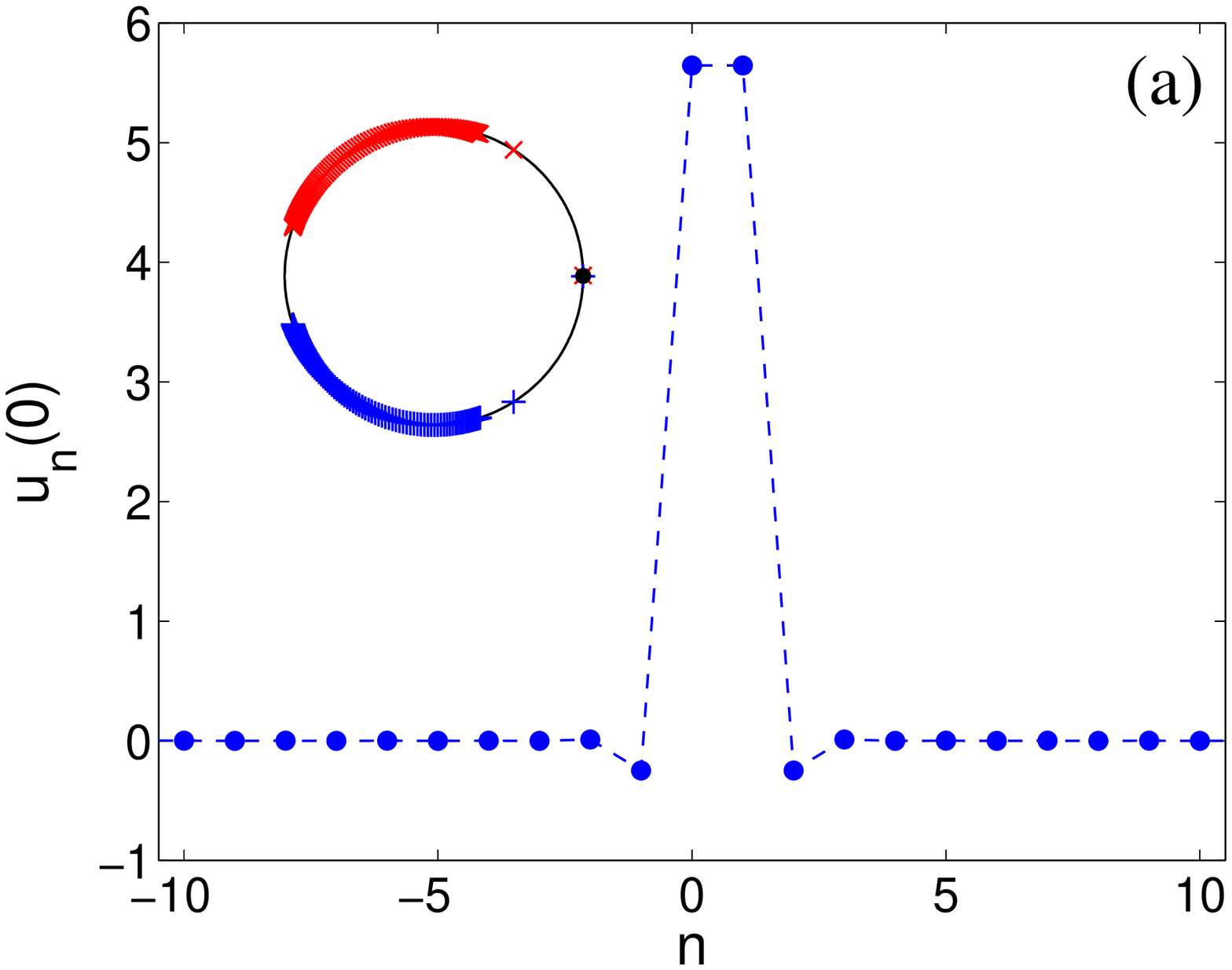} &
\includegraphics[width=5.cm]{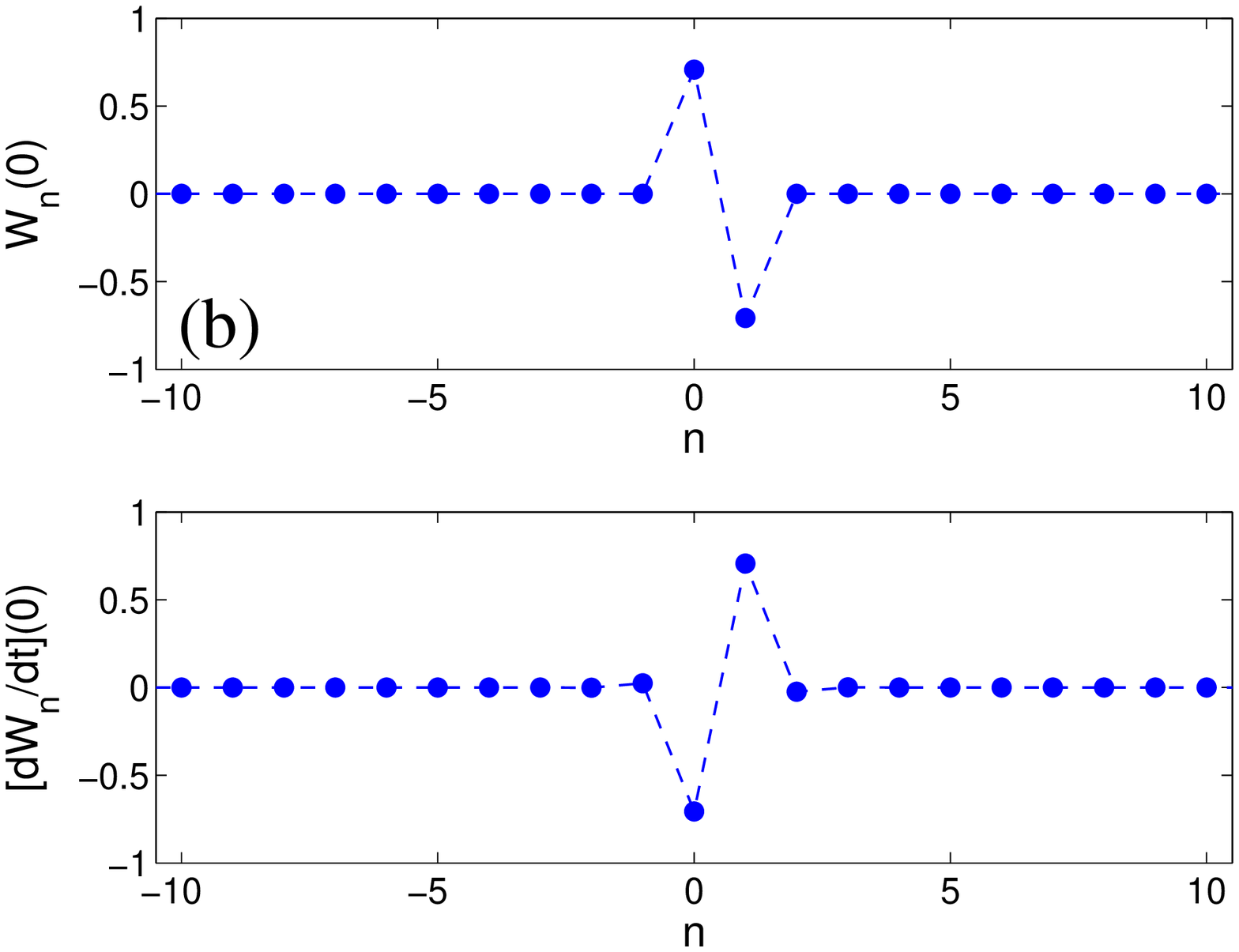} \\
\includegraphics[width=5.cm]{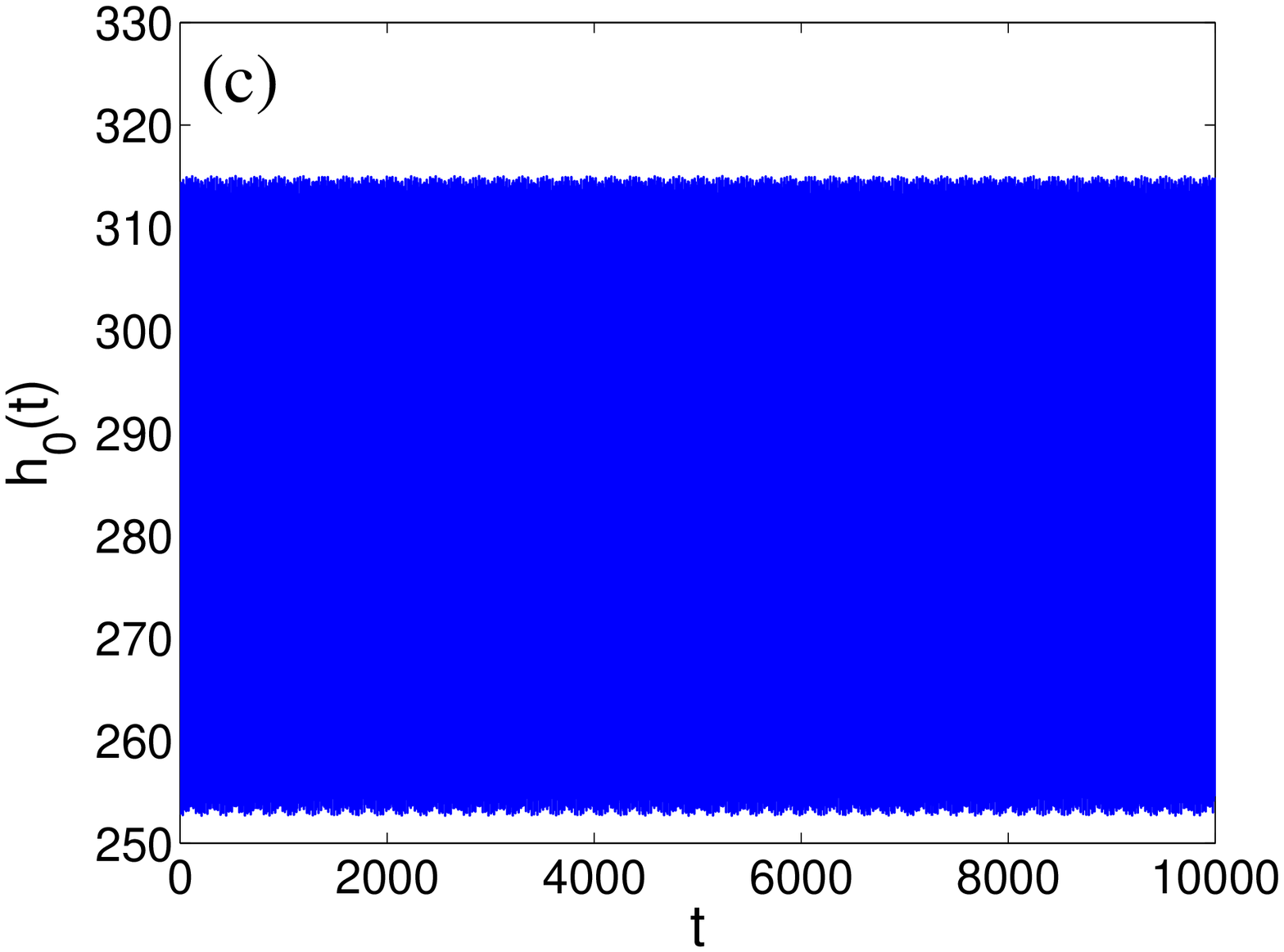} &
\includegraphics[width=5.cm]{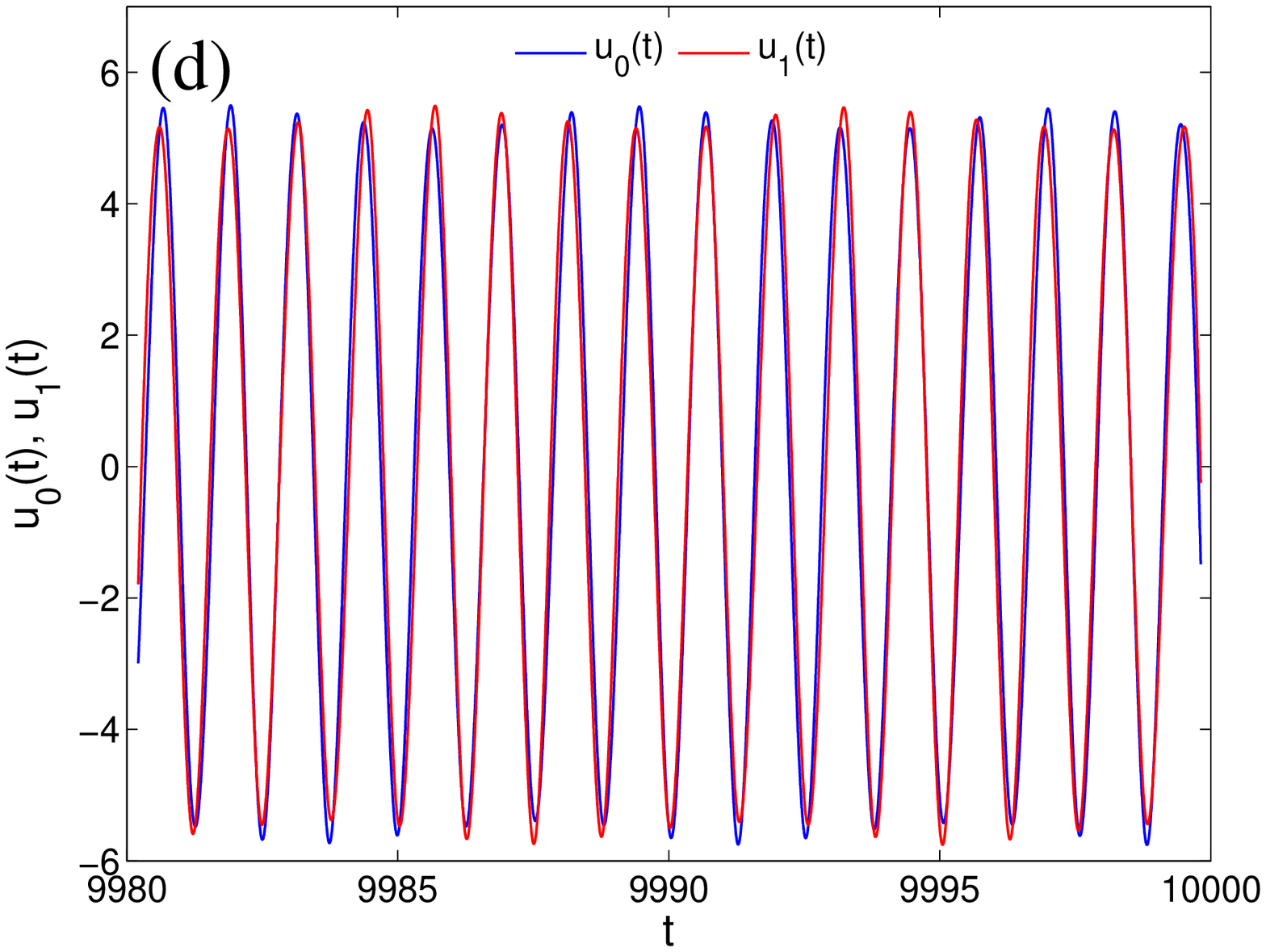} \\
\end{tabular}
\end{center}
\caption{Similar to Fig. \ref{fig:Morse1} but now for the
two-site in-phase breather in the hard $\phi^4$ potential for $\varepsilon = 1$ for $\omega_0 = 5$.}
\label{fig:quartic2}
\end{figure}

Figs.~\ref{fig:Morse2} and~\ref{fig:quartic2} show the
cases predicted to be nonlinearly stable
in the dynamics of two-site breathers.
For the soft Morse potential, we take the anti-phase breather for $\varepsilon = 0.04$
and $\omega_0 = 0.65$ ($T \approx 3.08 \pi$), with the perturbation parameter $\delta=0.025$.
The internal mode possesses the frequency $\Omega=0.1319$ and the nearest phonon band
is shifted by $m_0 = 2$ to $[0.22,0.3]$, so that $2 \Omega$ is inside the phonon band.
For the hard $\phi^4$ potential, we take the in-phase breather for $\varepsilon = 1$, $\delta=0.2$,
and $\omega_0 = 5$ ($T = 0.4 \pi$). The internal mode frequency is $\Omega=0.8013$
and the nearest phonon band is not shifted ($k_0 = 0$) and is located at $[1,2.23]$, so that $2 \Omega$ is again inside the phonon band.
In both cases, the internal mode and the phonon band in the upper half-circle have the same Krein signature.

Despite performing long-time dynamical simulations, an instability of the two-site breathers
does not seem to arise, confirming our theoretical predictions. Due to the
Hamiltonian nature of the dynamics and the
excitation of the initial perturbation, we observe a
quasi-periodic dynamics of the original two-site breathers, however,
there are fundamental differences observed between the oscillatory
dynamics of Figs.~\ref{fig:Morse2} and~\ref{fig:quartic2} and
the unstable ones involving growth (and the eventual destruction
of the two-site breather states) of Figs.~\ref{fig:Morse1}
and~\ref{fig:quartic1}.

%For the breather in the Morse potential
%on Fig. \ref{fig:Morse1}, a quasi-periodic breather forms with central sites oscillating in anti-phase, whose frequency is $\omega'_0=0.6347$ and the modulation frequency is $\omega_\mathrm{I}=0.1303$. The modulation frequency is close to the frequency $\Omega=0.1319$ of the internal mode, whereas the new breather frequency $\omega'_0$ is close to the perturbed breather frequency as $\omega'_0-\omega_0\approx0.015$. For the breather in the $\phi^4$ potential on Fig. \ref{fig:quartic2}, the quasi-periodic breather emerges with the new breather frequency $\omega'_0=5.0032$ and the modulation frequency $\omega_{I}=5.7960$. The modulation frequency is again close to the frequency of the internal mode $\Omega=0.8013$. The new breather frequency $\omega_0'$ is close to the perturbed breather frequency
%as $\omega'_0-\omega_0\approx0.003$. Hence we have the stable dynamics of the two-site considered in Figs. \ref {fig:Morse2} and \ref{fig:quartic2}.

\section{Conclusions and Future Challenges}

In the present work, we explored a broad question
regarding the
stability of multi-site breathers
in Klein--Gordon lattices. This question arises in a wide array of settings where discrete breathers
emerge, namely in nonlinear dynamical lattices.
Following the earlier work on discrete (and continuous)
nonlinear Schr{\"o}dinger
models~\cite{Cuccagna,KPS}, we systematically developed the
notion of the Krein signature for the wave
continuum and the internal modes associated with the
excited breather states. Subsequently, we produced asymptotic expansions
that illustrated the occurrence of the  nonlinear instability of multi-breathers,
when (second) harmonics of internal modes resonate with the wave continuum
of the opposite Krein signature.

However, we also revealed that contrary to the case of discrete (and continuous) nonlinear Schr{\"o}dinger
models, depending on the period of
the breather, it is possible to identify parametric regimes
for both soft and hard nonlinearities, where the multi-site breathers are nonlinearly stable.
We have confirmed all of the above conclusions through direct
numerical simulations of two on-site potentials, namely the
soft Morse and the hard $\phi^4$ potentials.

This instability poses a number of interesting questions for future
studies. On the one hand, it is relevant to explore theoretically
whether generalizations of this mechanism can be numerically observed
for higher-than-second harmonics (with a suitably modified growth
law).
Furthermore, the mechanism is expected to persist not only
for multi-site breather configurations bearing a larger
number of sites, but also in higher dimensional
setups and for structurally more complex excited state configurations,
such as vortex breathers.
On the other hand, the nonlinear nature of the instability and
its weaker-than-exponential growth poses a substantial
challenge towards its experimental realization.
Finally, a deeper understanding of the evolution stages of the
instability beyond its onset would be an important (albeit
potentially formidable) task for the dynamics of the nonlinear
models of interest.

\begin{acknowledgment}

\section*{Acknowledgments}
The authors are grateful to Dr. Avadh Saxena for seeding this collaboration
and for valuable discussions. P.G.K.~gratefully acknowledges the support of
NSF-DMS-1312856, as well as from the US-AFOSR under grant FA9550-12-1-0332,
and the ERC under FP7, Marie Curie Actions, People, International Research Staff
Exchange Scheme (IRSES-605096). P.G.K.'s work at Los Alamos is supported in part by the U.S. Department
of Energy. The work of D.P. is supported by the Ministry of Education
and Science of Russian Federation (the base part of the state task No. 2014/133).

\end{acknowledgment}

\end{document}